\documentclass[%
 reprint,
 amsmath,amssymb,
 aps,
]{revtex4-2}
\usepackage{xcolor}
\usepackage{graphicx}
\usepackage{dcolumn}
\usepackage{bm}

\begin{document}

\preprint{APS/123-QED}

\title{Optical control protocols for high-fidelity spin rotations of single SiV$^{-}$ and SnV$^{-}$ centers in diamond}

\author{Evangelia Takou}
\affiliation{Department of 
Physics, Virginia Polytechnic Institute and State University, 24061 Blacksburg, VA, USA}
\author{Sophia E. Economou}
\affiliation{Department of 
Physics, Virginia Polytechnic Institute and State University, 24061 Blacksburg, VA, USA}

\begin{abstract}
Silicon-vacancy and tin-vacancy defects in diamond are of interest as alternative qubits to the NV center due to their superior optical properties. While the availability of optical transitions in these defects is one of their assets, high-fidelity optical coherent control has not been demonstrated. 
Here, we design novel optical control schemes tailored to these defects. We evaluate the performance of arbitrary single-qubit rotations of the electron spin qubit both in the absence and presence of an external magnetic field, by taking into account both coherent and incoherent errors. We find that rotations in excess of $98.0\%$ ($T=4$~K) and $99.71\%$ ($T=6$~K) can be achieved for Si-V and Sn-V respectively in the presence of realistic relaxation and leakage errors.  
\end{abstract}

\maketitle

\section{Introduction}

Over the past decades, color centers in diamond have been investigated for their potential use as the hardware for solid-state quantum processing applications. The most well-known defect in a diamond host is the NV center. Its most prominent features are the long coherence times of the NV electron spin \cite{Wratchrup2009} and the room temperature operation \cite{Awschalom2005,Wrachtrup2005}. Practical experimental demonstrations regarding the NV center involve readout \cite{PhysRevLett.92.076401,Gupta:16,Lovchinsky836,MeteNV2019npj}, initialization \cite{Hanson2011,HansonNat2011}, entanglement generation schemes, as well as control of the surrounding nuclear bath \cite{Hanson2015,PhysRevLett.109.137602}. However, the optical control usually relies on the application of external perturbations such as strain/electric or magnetic fields, to lift the ground state degeneracy, or allow for spin-flipping transitions \cite{PhysRevX.6.041060,Awschalom2017}. In addition, NV centers have poor optical properties, with large phonon sideband and low probability of emission in the zero phonon line (ZPL). While this can be boosted using coupling to cavities and waveguides \cite{doi:10.1063/1.4948746}, the low ZPL emission still limits the performance of entanglement generation schemes. Moreover, the sensitivity of NV centers to charge noise introduces spectral diffusion to optical transitions \cite{PhysRevLett.110.027401}. 

As a result, alternative defects are explored in diamond for quantum information applications. Two that stand out are the negatively charged silicon vacancy (SiV$^-$) \cite{PhysRevLett.112.036405,MeteSiV2014NatCommun,Vuckovic2018,Vuckovic2018PRL,MeteSiV2018PRB,MeteSiV2018NatCommun,LukinSiV2018PRL,LukinSiVPRX2019,LukinSiV2020Nature} in diamond  and the newly emerging tin vacancy (SnV$^-$) color centers  \cite{PhysRevLett.124.023602,PhysRevB.99.205417,doi:10.1021/acs.nanolett.9b04495,VuckovicTin2021,Vuckovic2021b,VuckovicACSPhoton2020}. These spin $S=1/2$ systems, have excellent optical properties, such as narrow linewidths of the ZPL transition which comprises $70-80\%$ of the emitted light \cite{PhysRevB.89.235101}, small phonon sidebands, and spectral stability \cite{https://doi.org/10.1002/pssa.201700586, Rogers2014}. They belong to the $D_{3\text{d}}$ point group \cite{PhysRevLett.112.036405} and display inversion symmetry, which renders them robust to charge fluctuations, since, unlike the NV, they lack permanent electric dipole moment to first order \cite{PhysRevLett.77.3041}. Consequently, they are excellent indistinguishable single photon sources \cite{PhysRevLett.113.113602,PhysRevApplied.5.044010}, and they are immune to noise emerging from  integration into photonic devices \cite{PhysRevApplied.7.024031}. Furthermore, due to the large ground state splitting of the SnV$^-$ defect \cite{Aharonovich2019,PhysRevLett.124.023602}, no spin mixing is observed in the presence of external magnetic fields \cite{PhysRevLett.124.023602}, which can lead to improved optical control.
In addition, nuclear spin control has been achieved for SiV$^-$ in   nano-waveguides \cite{PhysRevB.100.165428, PhysRevLett.123.183602}. Initialization and readout of the SiV$^-$ have also been demonstrated in \cite{PhysRevLett.113.263602,PhysRevLett.119.223602}, while coherent control and single qubit rotations have been shown in \cite{MetePRL2014,Zhang:17,Becker2016,Mete2017,MeteSiVPRL2018}. In \cite{Mete2017}, the control of the electronic spin is achieved using MW pulses, resulting in slower rotations. This approach also requires microwave frequency generators and amplifiers, thus increasing the experimental demands. In \cite{Becker2016}, full SU(2) spin rotations were demonstrated via Ramsey interference generated by two temporally separated pulses. Each of these pulses originated from a single broadband laser that addressed both transitions of a $\Lambda$-system. To avoid driving unwanted transitions, a far off-resonant Raman pulse was used, which restricted the achievable rotation angles due to limitation of the laser power and did not mitigate the decoherence entirely due to the excitation of an unwanted excited state. Moreover, the fidelity of the rotations was not quantified nor were the error mechanisms investigated in detail. 

So far, the theoretical models of optical control usually involve approximations under which the off-resonant transitions are ignored. This is a good approximation for longer gate durations, i.e., narrowband pulses. For faster pulses, which are needed in these systems to ensure the control takes place well within the optical coherence and relaxation times, such off-resonant transitions cannot be ignored. Another issue typically present in defect systems is orbital mixing of the states caused by the Jahn-Teller effect or crystal strain. In the case of $\Lambda$-system schemes, each transition dipole couples to both driving fields, leading to
additional errors during the optical control.
Thus, these error mechanisms present in many defects give rise to the need for high-fidelity control techniques tailored to these systems.

In this paper, we address the aforementioned challenges by
developing all-optical control protocols for the SiV$^-$ and SnV$^-$ color centers in diamond. We start by demonstrating that existing approaches, in particular coherent population trapping, do not suffice for high-fidelity gates. We show how to eliminate cross-talk errors and reduce the number of unwanted leakage transitions by appropriately selecting the polarization of the lasers. We further optimize the gates by analyzing the full dynamics of the systems, and we identify the coherent errors as well as incoherent errors that arise due to unwanted excitations of the multi-level electronic structures. We resolve the leakage through two corrective methods, one available in the literature and one developed here, that allow for faster implementation of the rotations without compromising the gate fidelity.

This paper is organized as follows. In Sec.~\ref{Sec2} we describe the two main sources of errors of the optical control; the cross-talk and leakage. In Sec.~\ref{Sec3}, we discuss the cross-talk issue and provide an effective and simple solution based on polarization schemes. In Sec.~\ref{Sec4} we analyze two approaches for leakage mitigation. Finally, in Secs.~\ref{Sec5} and \ref{Sec6}  we apply our protocols to the SiV$^-$ and SnV$^-$ defects, respectively, and quantify the fidelity for various rotations angles and pulse durations.

\section{Overview of the problem and error mechanisms \label{Sec2}}

\begin{figure}[!htbp]
    \centering
    \includegraphics[scale=0.5]{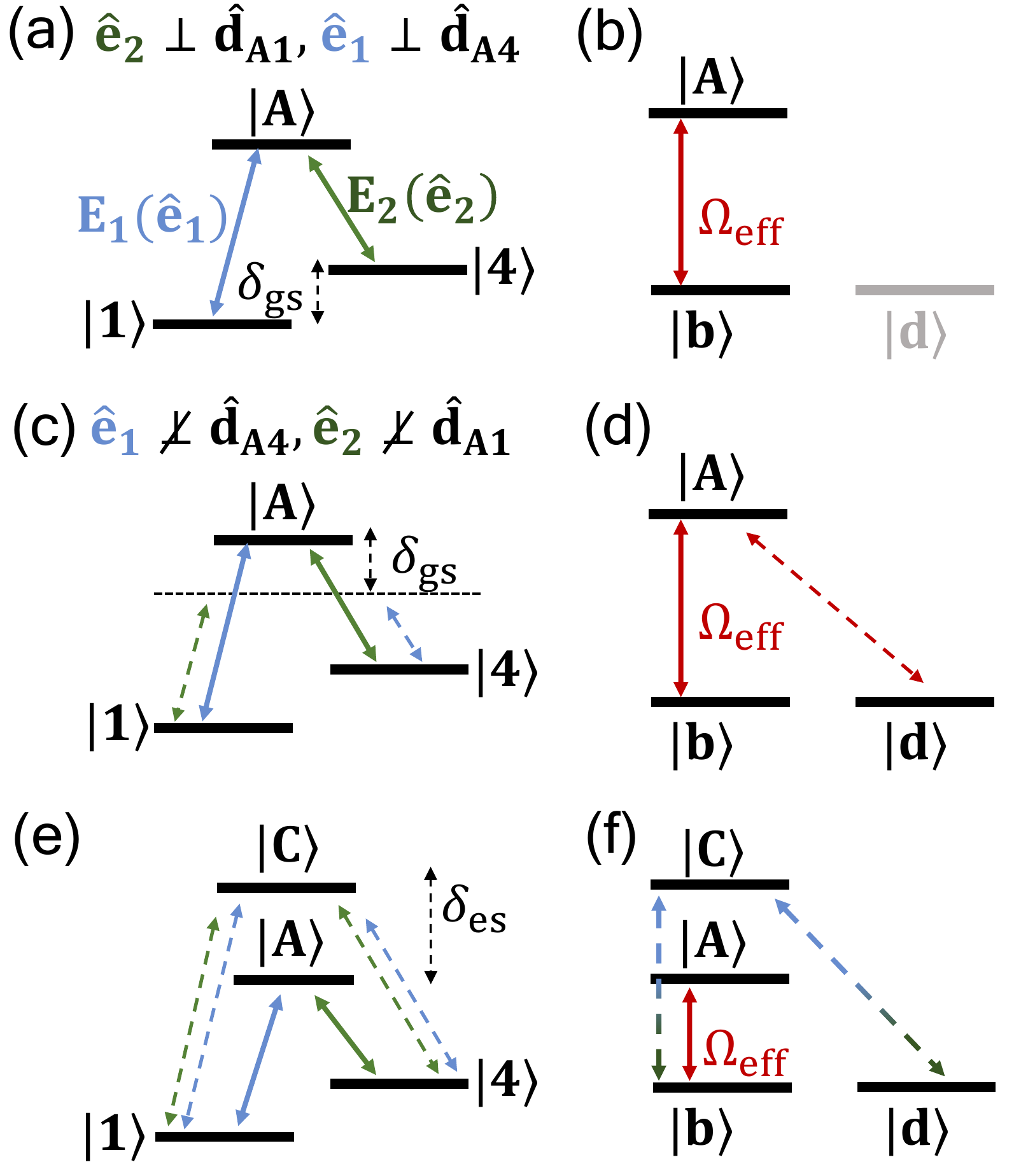}
    \caption{Summary of the error mechanisms for the SiV$^-$ system. (a) Ideal CPT scheme performed with two fields $E_1$ and $E_2$ acting on distinct transitions. The ground state splitting is denoted as $\delta_{\text{gs}}$. (b) Transformation to the dark-bright basis for the case of (a) successfully  decouples the dark state $|\text{d}\rangle$ and transitions are driven between the bright $|\text{b}\rangle$ and excited state $|\text{A}\rangle$. The effective Rabi frequency is expressed in terms of the Rabi frequencies in the lab frame, i.e. $\Omega_{\text{eff}}=\sqrt{|\Omega_1|^2+|\Omega_2|^2}$. (c) Cross-talk within the $\Lambda$-system leads to off-resonant errors (dashed green and blue arrows), that oscillate with an additional energy shift, $\delta_{\text{gs}}$. (d) In the presence of cross-talk couplings as shown in (c), the dark state is not completely decoupled. (e) Both laser fields drive the leakage transitions C1 and C4, introducing errors to the optical control. $\delta_{\text{es}}$ is the excited states splitting. (f) In addition to the cross-talk shown in (d), each laser drives the $|\text{b}\rangle\leftrightarrow |\text{C}\rangle$ and $|\text{d}\rangle\leftrightarrow |\text{C}\rangle$ transitions in the db-frame.}
    \label{fig:1}
\end{figure}

A well-known technique that provides fast all-optical control of $\Lambda$-systems is coherent population trapping (CPT). In CPT, the two transitions of a $\Lambda$-system are driven with two laser fields $E_1$ and $E_2$, each acting on a distinct transition, as shown in Fig.~\ref{fig:1}(a), which satisfy the two-photon resonance condition.
CPT is based on the destructive interference of the quantum processes driven by the different fields, which leads to trapping of the population into a dark state. In this so-called dark-bright frame, the dark state is completely decoupled from the dynamics of the other two levels in the system [Fig.~\ref{fig:1}(b)]. The transformation to the dark-bright frame defines the rotation axis of the qubit; by combining CPT with hyperbolic secant pulses,  we can design arbitrary single-qubit rotations, as explained in Appendix~\ref{SecApp0} and in Refs.~\cite{PhysRevB.74.205415,EconomouPRL2007}. 

In an ideal CPT scheme, the distinct couplings can be satisfied by either energy separation of the ground states, or polarization selection rules that ensure each transition is accessed by a single laser. However, energy separation alone does not guarantee negligible cross-talk errors for all gate durations, and the approximation of distinguishable couplings breaks down for broadband pulses. Unfortunately, the two transition dipoles are not orthogonal for the SiV$^-$ and the SnV$^-$ systems, leading to the cross talk (dashed arrows) shown in Fig.~\ref{fig:1}(c). The source of the cross talk and our solution to this problem will be explained in Sec.~\ref{Sec3}.  
For now, we stress that this setting is unavoidable if each laser field is chosen according to the polarization selection rules, i.e., such that its coupling to one of the two $\Lambda$-transitions is maximized. Henceforth, we refer to this approach as ``naive''.

In addition to the cross-talk, each laser field removes population from the $\Lambda$-subspace inducing thus leakage errors to the control. As an example, we show the leakage transitions of the SiV$^-$ system in Fig.~\ref{fig:1}(e), which occur with an off-resonant energy cost $\delta_{\text{es}}$. In the dark-bright frame, these errors translate into couplings between the dark/bright states and the unwanted excited level, $|\text{C}\rangle$ [Fig.~\ref{fig:1}(f)]. 

In the following sections, we propose schemes to resolve the cross-talk by polarization tuning of the lasers, as well as to counteract the leakage errors via pulse modulation. These protocols are analyzed in Sec.~\ref{Sec3} and Sec.~\ref{Sec4} respectively. The readers who are more interested in the numerical results could directly proceed with Sec.~\ref{Sec5} (for the SiV$^-$) and Sec.~\ref{Sec6} (for the SnV$^-$).

\section{Addressing Cross-talk errors \label{Sec3}}

\begin{figure}[!htbp]
    \centering
    \includegraphics[scale=0.39]{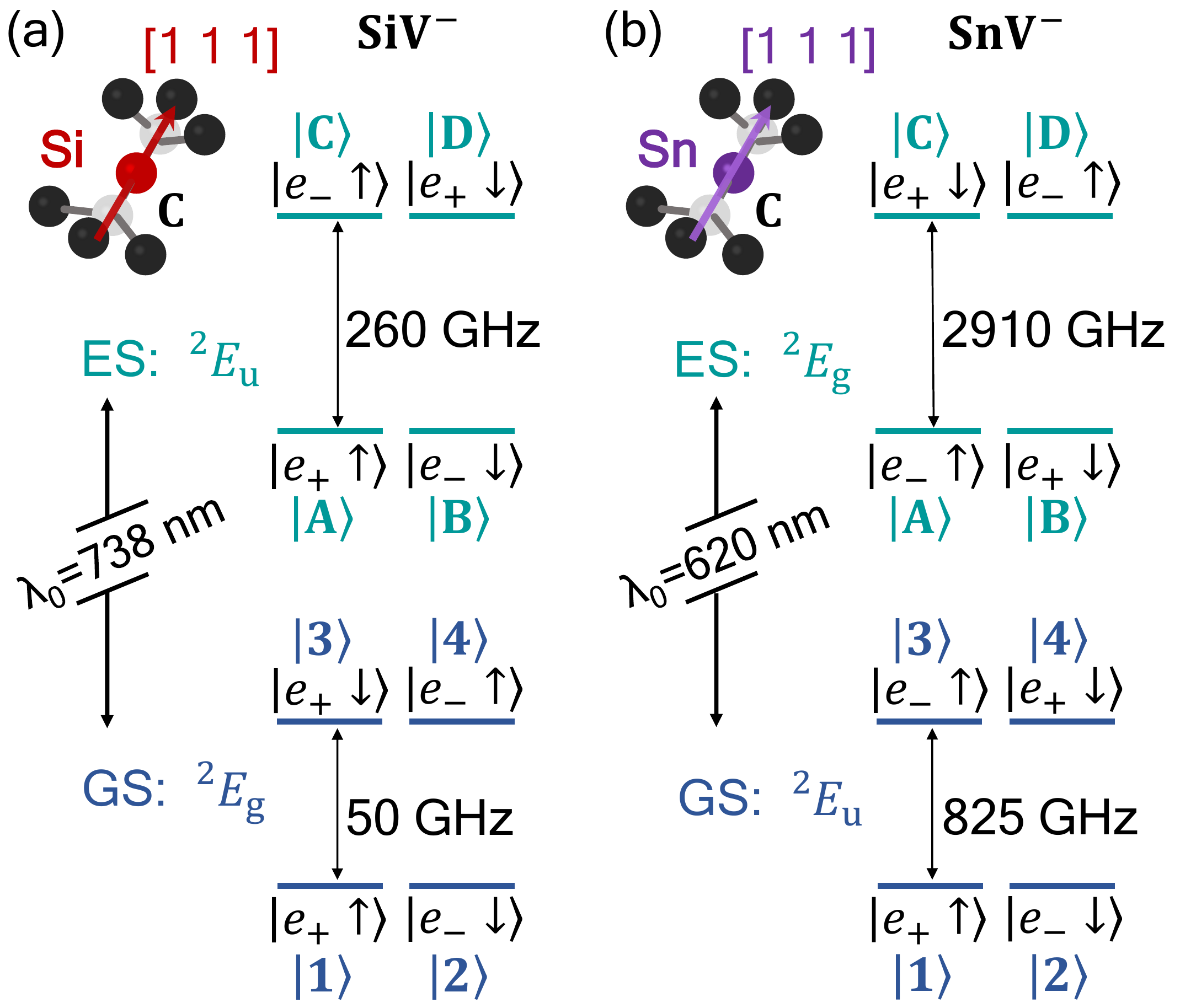}
    \caption{Electronic structure for the negatively silicon vacancy (SiV$^-$) (a), and for the negatively charged tin vacancy (SnV$^-$) in diamond (b), for zero magnetic fields. We label the states using the eigenstates of the spin-orbit coupling (i.e. $|e_{\pm}\rangle$) which is the largest perturbation term, but in the text, we use the notation $|1\rangle- |4\rangle$ for the ground states and $|\text{A}\rangle-|\text{D}\rangle$ for the excited states.  }
    \label{fig:2}
\end{figure}

One advantage of SiV$^-$ and SnV$^-$ defects is that $\Lambda$-schemes can be realized even at zero-magnetic fields, which simplifies the dynamics and facilitates experimental implementations of optical control. In Fig.~\ref{fig:2} we show the electronic structure for the SiV$^-$ and SnV$^-$ at zero magnetic fields; each ground- and excited-state manifold is pairwise degenerate. We follow the literature convention of labeling the ground states as $|1\rangle - |4\rangle$, and the excited states as $|\text{A}\rangle -|\text{D}\rangle$ (for more precise labeling we use the eigenstates of the spin-orbit coupling, $|e_\pm\rangle$, for the orbital part of the states). 

Based on group theory, the allowed optical transitions can be accessed by either linear $z$-polarization, which drives transitions between orbital states with the same symmetry, or by circular polarization, which drives transitions between the states $|e_{\text{g},\pm}\rangle \leftrightarrow |e_{\text{u},\mp} \rangle$ \cite{PhysRevLett.112.036405}. However, 
a small orbital mixing of the states caused by the Jahn-Teller effect \cite{PhysRevLett.112.036405} (or by crystal strain) introduces  non-zero $z$-dipoles ($x,y$ dipoles) to the transitions mainly accessed by $\sigma^\pm$ ($z$) polarization. Consequently, the choice of polarizations as dictated by selection rules would give rise to a cross-talk, i.e. coupling of each laser field to both $\Lambda$-transitions. In such a setting, the dark state is not completely decoupled from the dynamics, as shown in Fig.~\ref{fig:1}(d). 

The contribution of the off-resonant cross-talk is quantified by the ground state splitting of the qubit states. For the SnV$^-$ system, the errors average out more efficiently due to its large ground-state splitting ($\delta_{\text{gs}}^{\text{SnV$^-$}}=825$~GHz). These errors however are not negligible for SiV$^-$ (for which $\delta_{\text{gs}}^{\text{SiV$^-$}}=50$~GHz).
Nevertheless, when broadband pulses are considered, the cross-talk becomes the central source of infidelity for both defects since off-resonant transitions are more strongly coupled.

We propose a simple cross-talk elimination scheme that is achieved by tuning the laser field polarizations. We consider as an example the SiV$^-$ system and express the direction and polarization of the laser fields in the defect coordinate frame. In an experimental setup, most diamond samples are cut with $[0~0~1]$ as the surface normal. Thus, the polarization directions that we express in the internal coordinate frame would require a non-zero angle of incidence on the sample.

We assume two lasers of $xz$-polarization and $y$-propagation, where the former drives the A1 transition and the latter the A4 transition. To fix the orthogonality of each laser with one of the $\Lambda$-transitions, we require that its polarization is orthogonal to the corresponding dipole, $\textbf{d}_{ij}$. In this particular case of $\Lambda$-system, we require $\textbf{d}_{\text{A1}}\cdot \textbf{E}_2=0$ and $\textbf{d}_{\text{A4}}\cdot \textbf{E}_1=0$, from which we find that the electric fields need to be defined as:
\begin{equation}\label{Eq4}
    \textbf{E}_1=E_{01}\left(\hat{\textbf{x}}-\frac{\langle  p_x \rangle_{\text{A4}}}{\langle p_z \rangle_{\text{A4}}}\hat{\textbf{z}}\right)e^{i(k_1y-\omega_1 t)}+\text{cc.},
\end{equation}
\begin{equation}\label{Eq5}
    \textbf{E}_2=E_{02}\left(\hat{\textbf{x}}-\frac{\langle  p_x \rangle_{\text{A1}}}{\langle p_z \rangle_{\text{A1}}}\hat{\textbf{z}}\right)e^{i(k_2y-\omega_2 t)}+\text{cc.}.
\end{equation}
Here $\langle p_k\rangle_{ij}=\langle \psi_i |p_k|\psi_j\rangle$ (with $k\in\{x,y,z\}$) is the transition dipole overlap that can be calculated according to group theory. The polarizations of the lasers that satisfy the orthogonality conditions are not unique; we choose to restrict the polarization vectors in the $xz$ plane, in which case the polarizations are uniquely determined. The definitions of Eq.~(\ref{Eq4}) and Eq.~(\ref{Eq5}) can be generalized easily to other choices of $\Lambda$-systems or other polarization directions. For the SnV$^-$ system, we chose $yz$-polarization instead, and the reasons for this choice are explained in Sec.~\ref{Sec6a} and in Appendix \ref{SecApp5}.  

Throughout the paper, we combine the sech-based CPT scheme with \textbf{E}-field polarizations that satisfy the orthogonality conditions to design arbitrary gates free from cross-talk errors.

\section{Corrective methods for leakage suppression \label{Sec4}}

\subsection{General strategy of Magnus expansion \label{Sec4a}}

As we mentioned in Sec.~\ref{Sec3}, we resolve the cross-talk issue of the $\Lambda$-system by redefining the polarization of the laser fields. 
However, leakage errors reduce the gate fidelity of fast optical control schemes. To counteract this problem, we use a Magnus-based expansion approach developed in Ref.~\cite{Clerk2021}. Here we outline the basic steps of the method, and we provide further details about the procedure we follow in Appendix~\ref{SecApp5}. 

Let us consider a generic Hamiltonian $H(t)$ given by:
\begin{equation}\label{Eq10}
    H(t)=H_0(t)+\epsilon V(t),
\end{equation}
where $H_0(t)$ implements our analytically solvable target gate, and $V(t)$ introduces an error to the dynamics generated by $H_0(t)$. The error term is assumed to be perturbative, as it contains off-resonant terms, oscillating faster than $H_0$. To mitigate these errors, we additionally consider a control Hamiltonian $W(t)$, such that the total Hamiltonian is modified into
\begin{equation}\label{Eq11}
    \bar{H}(t)=H_0(t)+\epsilon V(t)+W(t).
\end{equation}
Further, the control Hamiltonian is expanded in a power series according to:
\begin{equation}\label{Eq12}
    W(t)=\sum_{k=0}^\infty \epsilon^k W^{(k)}(t).
\end{equation}
By going into the interaction picture of $H_0(t)$, the Hamiltonian transforms into $\bar{H}_\text{I}(t)=\epsilon V_\text{I}(t)+W_\text{I}(t)$, and the total evolution operator becomes
\begin{equation}\label{Eq14}
    U(t)=U_0(t)U_{\text{I}}(t),
\end{equation}
where $U_0$ is the ideal gate, and $U_\text{I}(t)$ is generated by the error and control Hamiltonian. The implementation of the ideal gate is achieved if $U_{\text{I}}(T)=\textbf{1}$ (where $T$ is the gate time), such that $U(T)=U_0(T)$, based on Eq.~(\ref{Eq14}). To this end, the evolution operator $U_\text{I}(t)$ is expanded in a Magnus series, and as was shown in Ref.~\cite{Clerk2021}, the solutions for the control are obtained iteratively via the equation:
\begin{equation}\label{Eq16}
    \epsilon^n \int_0^{T}dt' W_{\text{I}}^{(n)}(t')=-i\sum_{k=1}^n\Omega_k^{(n-1)}(T),
\end{equation}
where $\Omega_k$ is the $k$-th Magnus expansion order. In this work, we focus on first order corrections, i.e. we truncate the Magnus series to the first order, which leads to the equation:

\begin{equation}\label{Eq18}
    \epsilon\int_0^{T}dtV_{\text{I}}(t)=-\int_0^{T}dtW_\text{I}^{(1)}(t).
\end{equation}
Equation~(\ref{Eq18}) can be reformulated into a linear system of equations via the decomposition of the error and control part into an operator basis, which enables to rewrite it as \cite{Clerk2021}:
\begin{equation}\label{Eq19}
    B\textbf{x}^{(1)}=\textbf{y}^{(1)},
\end{equation}
where $B$ is a matrix that encodes the dynamics of $H_0$, $\textbf{y}^{(1)}$ are the error terms, and $\textbf{x}^{(1)}$, is a vector that contains the solutions to the first order of control expansion. 

An essential requirement of the Magnus scheme is that the control Hamiltonian is decomposed to at least the same operators as the errors, in the final interaction frame of $H_0$. However, it is not a strict requirement that the control pulse has access to all error transitions in the initial frame. For both defect systems, the leakage transitions that remove the population outside of the $\Lambda$-subspace correspond to the C transitions. To cancel out the leakage in both cases, we need to modify only one of the original sech pulses that drive the $\Lambda$-transitions.
 
As we already mentioned, the control Hamiltonian is expanded in a power series. In our case, we consider the total envelope:

\begin{equation}
    g^{(n)}(t)=g_1^{(n)}(t)\cos(\omega_\text{d} t)+g_2^{(n)}(t)\sin(\omega_\text{d} t),
\end{equation}
composed of two $\pi/2$-shifted envelopes $g_{l}^{(n)}(t)$, which are expanded in Fourier series. In particular, we use only the cosine terms with $g_l^{(n)}(t)$ given by:
\begin{equation}
    g_l^{(n)}(t)=\sum_k c_{l,k}^{(n)} \left(1-\cos\left(\frac{2\pi t k}{T}\right) \right),
\end{equation}
where $n$ denotes the Magnus expansion order and $k$ denotes the Fourier expansion order. We have also fixed $g_l^{(n)}(0)=g_l^{(n)}(T)=0$ such that the corrective pulse is zero at the beginning and end of the evolution. Throughout this paper, we always truncate the Magnus expansion to the first order, i.e. we set $n=1$.

The driving frequency of the control $\omega_\text{d}$ is another free parameter that can be tuned to lead to the most effective leakage cancellation. Nonetheless, introducing and modifying a new laser field is more challenging experimentally. To that end, we restrict the control pulse to have the same frequency with the original pulse that we modulate. 

\subsection{DRAG framework \label{Sec4b}}

An alternative route to leakage suppression is based on the adiabatic removal of errors, which we analyze in this subsection. The DRAG technique is a widely known method, extensively used for correcting leakage errors in superconducting qubits \cite{PhysRevA.83.012308,DRAG10yrs,PhysRevLett.103.110501}. Based on the DRAG formalism, analytically derived controls are obtained via a time-dependent Schrieffer-Wolff transformation. 
The generator of the transformation is $A(t)=e^{-iS(t)}$, where $S(t)$ is a Hermitian operator, and leads to the effective DRAG Hamiltonian
\begin{equation}\label{Eq24}
    H_\text{D}=A^\dagger H_{\text{db}} A +i \dot{A}^\dagger A.
\end{equation}
The dark-bright frame Hamiltonian is given by:

\begin{equation}
    H_{\text{db}}=(\Omega_{\text{eff}}f(t)\sigma_{\text{be}}+\text{H.c.})-\Delta\sigma_{\text{ee}},
\end{equation}
where $\Delta$ is the two-photon detuning and $f(t)=\text{sech}(\sigma(t-t_0))$. Also, $|\text{b}\rangle$ is the bright state and $|\text{e}\rangle$ the excited state. By requiring that the frame transformation vanishes at the boundaries (i.e. $A(0)=A(T)=\textbf{1}$), the target evolution in the initial (in this case, dark-bright) and DRAG frames remains the same at the end of the pulse. To reduce the leakage errors, one needs to find an appropriate adiabatic transformation, $S(t)$ that respects the boundary conditions. Besides this restriction, $S(t)$ can be an arbitrary Hermitian operator, which allows for the suppression of leakage errors. 

The original DRAG scheme is designed to cancel out leakage errors of a ladder-type system 
(e.g. transmon), which are caused by transitions between consecutive levels.
In our work, we extend this formalism to a $\Lambda$-system. This is qualitatively different, since 
in our case the population is removed from the system via transitions that link the
ground (qubit) states to an unwanted excited level. Moreover, the complexity is increased, since
each leakage transition is driven by both laser fields used for the CPT control.

In the DRAG framework, $H_\text{D}$ has to be constrained in a way that it implements an ideal evolution dictated by a target Hamiltonian. In our case, the target Hamiltonian as defined in the CPT frame has the form:
\begin{equation}\label{Eq25}
    H_{\text{target}}=\frac{h_x^{(0)}(t)}{2}\sigma_{x,\text{be}}+h_z^{(0)}(\sigma_{\text{bb}}-\sigma_{\text{ee}}),
\end{equation}
where $|\text{d}\rangle$ ($|\text{b}\rangle$) is the dark (bright) state, $|\text{e}\rangle$ is the excited state of the $\Lambda$-system ($|\text{e}\rangle=|\text{A}\rangle$), and $h_z^{(0)}$ is the two-photon detuning. Also, we have defined $\sigma_{x,ij}=|i\rangle\langle j| +|j\rangle \langle i|$ and $\sigma_{y,ij}=-i(|i\rangle\langle j|-|j\rangle\langle i|)$. At this point, we should emphasize that our treatment is different from Ref.~\cite{PhysRevA.83.012308}, where the leakage-robust gates are designed according to a target qubit-Hamiltonian in the rotating frame. Instead, to reduce the leakage errors from the qubit subspace ($|\text{d}\rangle$, $|\text{b}\rangle$), we formulate an indirect treatment which involves the bright-excited subspace.

Based on the target Hamiltonian of Eq.~(\ref{Eq25}), our target constraints are:
\begin{eqnarray}
h_x^{(n)}&=&\text{Tr}[H_\text{D}^{(n)}\sigma_{x,\text{bA}}],\\
h_y^{(n)}&=&\text{Tr}[H_\text{D}^{(n)}\sigma_{y,\text{bA}}]=0,\\
h_z^{(n)}&=&\text{Tr}[H_\text{D}^{(n)}(\sigma_{\text{bb}}-\sigma_{\text{AA}})].
\end{eqnarray}
The zero-th order target constraints ensure that $H_\text{D}^{(0)}= H_{\text{target}}$. To satisfy the decoupling of the $\Lambda$-system from the $|\text{C}\rangle$ leakage subspace we require the following decoupling constraints, with $k\in\{x,y\}$:
\begin{eqnarray}
\text{Tr}[H_\text{D}^{(n)}\sigma_{k,\text{dC}}]&=& 0,\\
\text{Tr}[H_\text{D}^{(n)}\sigma_{k,\text{bC}}]&=& 0,\\
\text{Tr}[H_\text{D}^{(n)}\sigma_{k,\text{AC}}]&=& 0,
\end{eqnarray}
as well as:
\begin{equation}
\text{Tr}[H_\text{D}^{(n)}\sigma_{k,\text{dA}}]= 0,
\end{equation}
which ensures that in the DRAG frame there is no transition between the dark and excited states.  Intuitively, for any order of $H_\text{D}^{(n)}$ with $n\geq 0$, the elements of the DRAG Hamiltonian that do not correspond to the target subspace should be zero. To obtain the $n$-th order DRAG Hamiltonian, we expand both $S(t)$ and $H_{\text{db}}$ in power series. The appropriate pulse modifications to the initial fields are obtained by satisfying the constraints consistently. In the particular case of $R_x(\pi)$ rotations (where the two-photon detuning is zero), the modulation of one laser vanishes, which in terms of experimental requirements matches the Magnus scheme. More details about the analytic derivation of the corrective envelopes are provided in Appendix \ref{SecApp6}.

\section{Control of SiV$^-$ system \label{Sec5}}

\subsection{Zero magnetic fields \label{Sec5a}}

\begin{figure}[!htbp]
    \centering
    \includegraphics[scale=0.55]{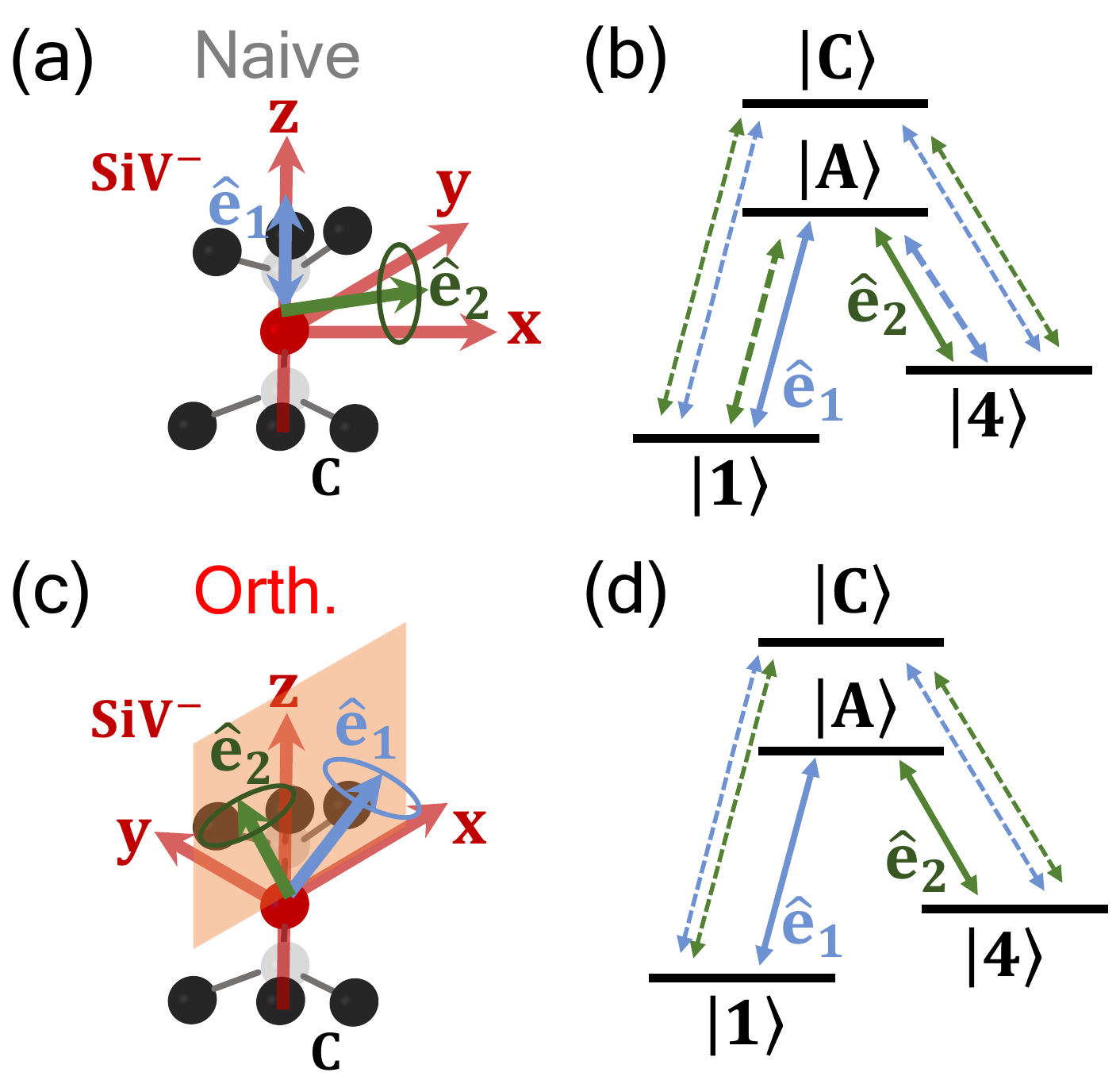}
    \caption{The selection rules-based \textbf{E}-field polarizations of (a) lead to cross-talk and four leakage transitions shown in (b). (c,d) The redefined polarizations in the $xz$-plane can eliminate the cross-talk of (b).}
    \label{fig:3}
\end{figure}

We begin by testing our protocols for the SiV$^-$ system at $B=0$~T. The bright transitions at zero magnetic fields are the A1, C1, B2, D2, B3, D3, A4 and C4 transitions. We consider the $\Lambda$-system formed by the states $|1\rangle$, $|4\rangle$ and $|\text{A}\rangle$ shown in Fig.~\ref{fig:3}(b). By choosing the $|\text{A}\rangle$ state to be the excited state of our $\Lambda$-system, we avoid downward orbital relaxations (which are more likely than the upward ones) that are present in the higher excited manifold. 

\begin{figure*}[!htbp]
    \centering
    \includegraphics[scale=0.52]{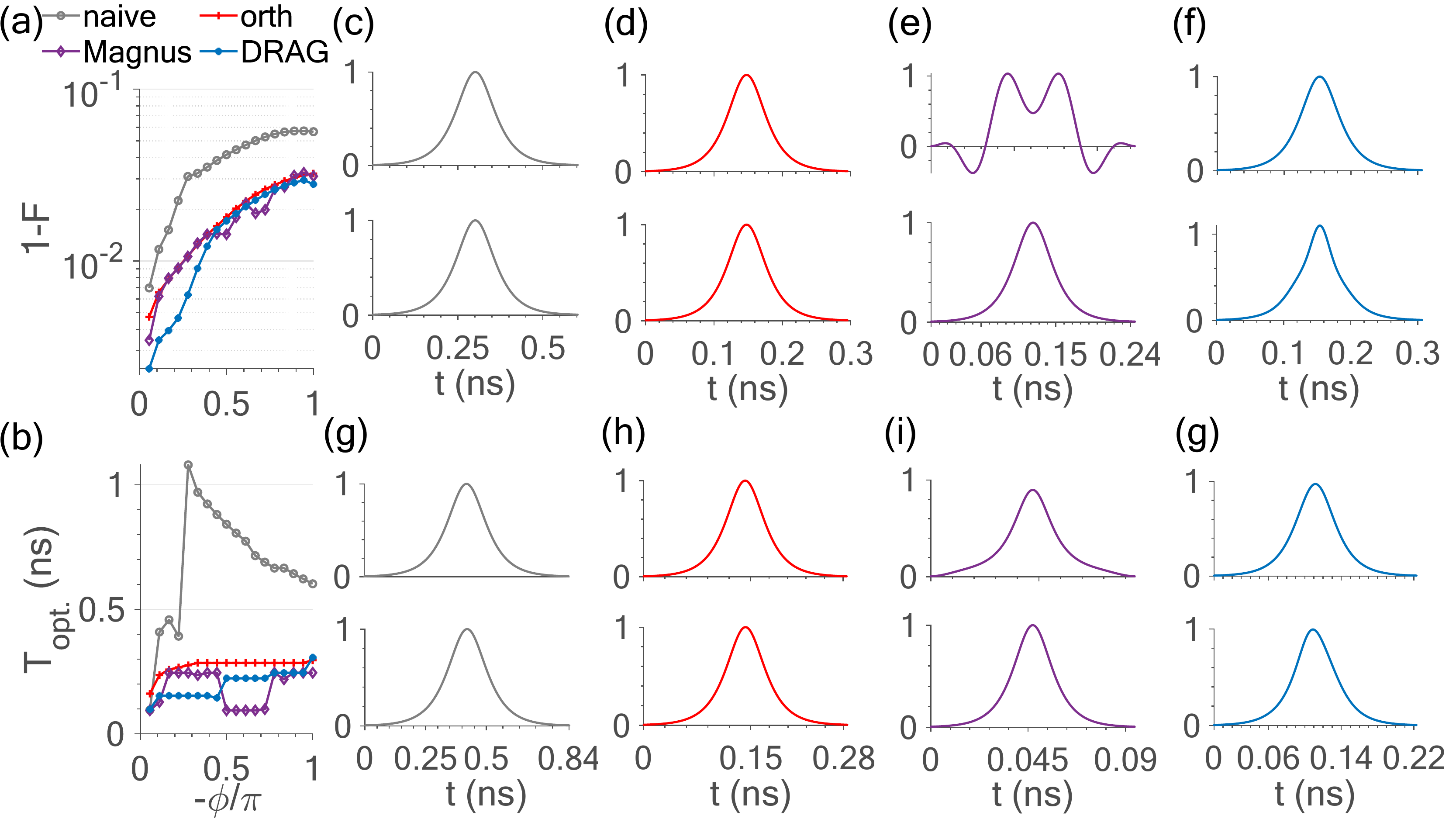}
    \caption{Optical control of SiV$^-$ at $B=0$: Infidelity of $R_x(\phi)$ (a), and optimal gate time (b), corresponding to four different protocols. The pulse envelopes for $R_x(-\pi)$ rotations are shown in (c), (d), (e), (f) and the pulse envelopes for $R_x(-\pi/2)$ are depicted in (g), (h), (i), (g). Gray: naive, red: orthogonal, purple: Magnus and blue: DRAG.  }
    \label{fig:4}
\end{figure*}
We assume no initialization errors and a temperature of $T=4$~K. At this temperature according to Ref.~\cite{PhysRevLett.113.263602}, the spin relaxation time is $T_{1,\text{spin}}=2.4$~ms, the orbital relaxation is $T_{1,\text{orbit}}=38$~ns, and the dephasing time $T_2^*=35$~ns. We also set the optical lifetime to be $\tau=4.5$~ns \cite{Becker2016}.

We start from the simplest approach of controlling the electronic spin, which we refer to as naive, which is simply based on the CPT control of an ideal Lambda system. It utilizes two laser fields according to the polarization selection rules; in this setup the A1 transition is driven by $z$-polarized light, and the A4 transition by circularly polarized light [see Figs.~\ref{fig:3}(a), (b)]. Notice that we define the polarizations in the defect coordinate frame. This choice of polarizations is far from optimal, since it introduces many off-resonant errors. In particular, in Fig.~\ref{fig:3}(b) we show all the possible transitions driven by the naive laser fields. The main error is the cross-talk within the $\Lambda$-system, as this involves a transition that is detuned by only $\delta_{\text{gs}}=50$~GHz, which is the ground state splitting. Each laser field additionally drives transitions C1 and C4 (off-resonant by $\delta_{\text{es}}=260$~GHz), introducing leakage outside of the $\Lambda$-subspace. 

In the naive approach, the leakage issue can be partially resolved by considering more narrowband pulses. Nevertheless, the relaxation mechanisms for longer pulses are detrimental to the optical control of the SiV$^-$, leading to enhanced gate errors. In Fig.~\ref{fig:4}(a), we show (gray curve) the infidelity of arbitrary rotations, $R_x(\phi)$, and in Fig.~\ref{fig:4}(b) the optimal gate time. The naive approach is the slowest of all our proposed protocols, since it balances the trade-off between leakage and cross-talk errors with relaxation-induced effects. The sech pulses for the optimal implementation of $R_x(-\pi)$ and $R_x(-\pi/2)$ by the naive approach, are shown in the panels of Fig.~\ref{fig:4}(c) and Fig.~\ref{fig:4}(g) respectively.

As mentioned in Sec.~\ref{Sec3}, the cross-talk can be completely eliminated by redefining the polarization of the two driving fields. In particular, we choose the polarization vectors to be in the $xz$-plane as shown in Fig.~\ref{fig:3}(c). We refer to this approach as orthogonal, shown with the red curve in Fig.~\ref{fig:4}. This protocol allows us to use more broadband pulses (still well-protected from leakage errors), while simultaneously reducing the effect of relaxations.  
Consequently, the pulses in the orthogonal scheme can be up to four times faster compared to the naive approach [see Fig.~\ref{fig:4}(b)], and the rotations display lower infidelity [see Fig.~\ref{fig:4}(a)]. In Fig.~\ref{fig:4}(d) and Fig.~\ref{fig:4}(h) we show the pulse envelopes of the orthogonal method for the optimal implementation of $R_x(-\pi)$ and $R_x(-\pi/2)$ rotations, respectively. 

The orthogonal approach sets the upper bound to the gate fidelities in the absence of additional pulse shaping. However, the optimal gate time of this method still lies within the regime where relaxation errors have non-zero contribution. It is also apparent that by reducing the gate time, the leakage errors gradually become more important. To mitigate the unwanted couplings to the upper excited manifold we use the corrective techniques we described in Sec.~\ref{Sec4}, and combine them with the orthogonal scheme to avoid the cross-talk within the $\Lambda$-system. To minimize the experimental overhead, we do not introduce new pulse envelopes, but instead modify the initial laser fields.

The first corrective method that we employ is the Magnus-based scheme \cite{Clerk2021}. In our protocol, we modify only one of the initial laser fields, which in this case is the laser field $E_1$ that drives the A1 transition, while the laser field $E_2$ that drives the A4 transition remains intact. Both initial fields introduce leakage outside of the $\Lambda$-system, via the excitation of the C1 and C4 transitions [each driven by both lasers as shown in Fig.~\ref{fig:3}(d)]. The modulated pulse has additional cosine envelopes, and the Fourier coefficients are obtained by solving a linear system of equations that we describe in Appendix~\ref{SecApp5}. To optimize the performance of the Magnus scheme, we search solutions that reduce the gate error for different Fourier series truncation and gate time intervals, while keeping the Magnus truncation to the first order.

We show the infidelity of the Magnus protocol in Fig.~\ref{fig:4}(a), and the optimal gate time in Fig.~\ref{fig:4}(b) (purple curve). The Magnus scheme allows us to reach an even faster regime while simultaneously restricting the leakage errors contributions. Therefore, with a simple modulation of one of the initial pulses we can retain the same fidelity as in the orthogonal scheme. The optimal pulse envelopes for $R_x(\pi)$ and $R_x(-\pi/2)$ are shown in Fig.~\ref{fig:4}(e) and Fig.~\ref{fig:4}(i) respectively. In both cases, the top panel corresponds to the modified pulse. 

The alternative corrective method for leakage suppression is the DRAG technique. This scheme requires pulse modulation of both initial laser fields, but in the particular case of $R_x(\pi)$ rotations, the correction to the field driving the A1 transition goes to zero. This is a consequence of the condition for achieving $R_x(\pi)$ gates, which requires zero two-photon detuning, leading to vanishing correction for one pulse. We notice that for rotation angles $\phi>-\pi/2$ the DRAG method (blue curve) has a longer optimal gate time 
compared to the Magnus method. For rotation angles $\phi<-\pi/2$, however, the gate time is further reduced [Fig.~\ref{fig:4}(b)] and the fidelity enhancement becomes more apparent [see blue curve of Fig.~\ref{fig:4}(a)]. We should also mention that the DRAG pulse modulations are obtained analytically [see Appendix~\ref{SecApp6}], but we also perform a simple optimization by redefining the amplitude of the corrections. The optimal pulses for $R_x(\pi)$ and $R_x(-\pi/2)$ are displayed in Fig.~\ref{fig:4}(f) and Fig.~\ref{fig:4}(g) respectively.

\subsection{Non-zero magnetic fields \label{Sec5b}}

In the presence of an external non-axial $\textbf{B}$-field, the pairwise degeneracy of each manifold is lifted, and all transitions become allowed and are no longer spin-conserving. This phenomenon is caused by the off-axial Zeeman interaction that gives rise to $S_xB_x$ and $S_yB_y$ terms, which cause spin-mixing of the states.

Arbitrary magnetic field directions are more difficult to implement experimentally since a vector magnet is required. For this reason, we assume a fixed magnetic field orientation where the $B_j$ magnetic field components in the SiV$^-$ frame are expressed in terms of the $B_\parallel$ and $B_\perp$ magnetic field strengths in the lab frame. The lab frame magnetic fields in an experimental setting would be applied parallel and perpendicular to the cryostat axis, where the sample is placed. The parallel magnetic field strengths reach up to $|B_\parallel|=9$~T and the perpendicular up to $|B_\perp|=3$~T. In the coordinate frame of the SiV$^-$ defect, we define the magnetic fields as:
\begin{equation}\label{Eq7}
    B_x=B_\parallel \cos (54.7^o)+B_\perp \sin(54.7^o),
\end{equation}
\begin{equation}\label{Eq8}
    B_y=0,
\end{equation}
and
\begin{equation}\label{Eq9}
    B_z=B_\parallel \sin (54.7^o)-B_\perp \cos(54.7^o),
\end{equation}
where $\gamma=54.7~^o$ is the angle between the symmetry axis [1~1~1] and the $(1~0~0)$ sample surface. 

The spin coherence of the $S=1/2$ systems shows an angular dependence on the direction of the external magnetic field. Larger deviation from the symmetry axis results in enhanced spin-mixing, which consequently reduces the spin coherence. In particular, according to Ref.~\cite{PhysRevLett.113.263602}, the reported $T_{1,\text{spin}}$ for the SiV$^-$ reduces to $3.6~\mu$s at $20^o$ misaligned field and to 60~ns at $70^o$ misalignment. In our simulations, we assume a spin relaxation time of $T_{1,\text{spin}}=300$~ns for the SiV$^-$.

We consider zero two-photon detuning corresponding to $R_x(\pi)$ gates, and we examine the $\Lambda$-system formed by the states $|1\rangle$, $|2\rangle$ and $|\text{A}\rangle$. We study only the performance of the orthogonal method and using the results of the analysis of Sec.~\ref{Sec5a}, we assume a fixed laser power, that balances the contribution of relaxations and leakage errors. 

\begin{figure}[!htbp]
    \centering
    \includegraphics[scale=0.55]{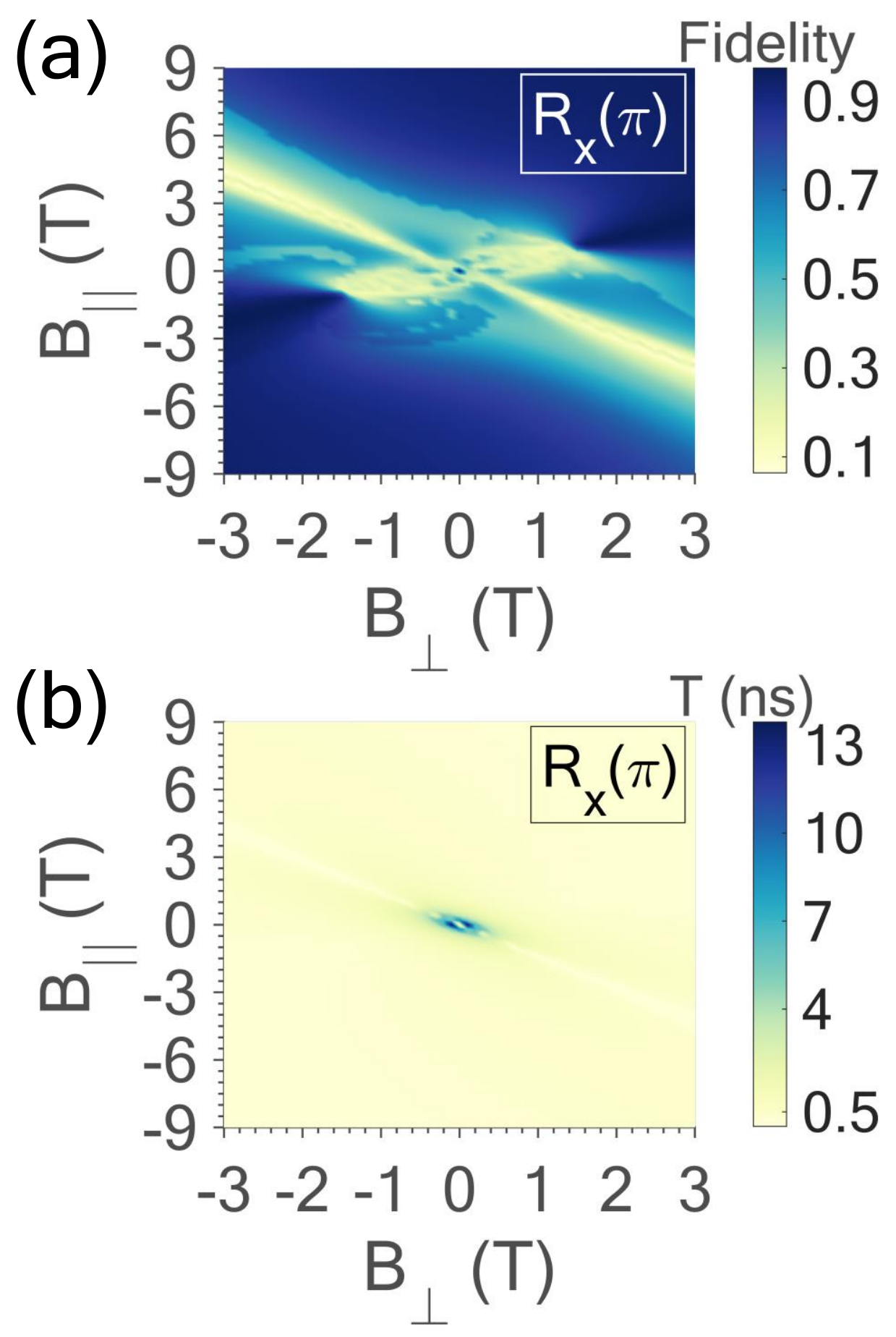}
    \caption{Optical control of SiV$^-$ at $B\neq 0$: Dependence of the fidelity (a) and duration (b) of $R_x(\pi)$ gates on the parallel and perpendicular (with respect to the cryostat axis) magnetic field strengths. For $B\neq0$ we consider the A1-A2 $\Lambda$-system. The regions of low fidelity correspond to weakly excited $\Lambda$-system.}
    \label{fig:5}
\end{figure}

In Fig.~\ref{fig:5}, we show the fidelity [Fig.~\ref{fig:5}(a)] and gate time [Fig.~\ref{fig:5}(b)] of $R_x(\pi)$ gates for the SiV$^-$ for a fixed laser field intensity. The maximum fidelity for the SiV$^-$ corresponds to $F_{\text{max}}^{\text{SiV$^-$}}=0.975$ for $B_\perp=-2.8$~T and $B_\parallel=-2$~T. The corresponding gate time is $T=0.235$~ns (versus $T=0.3$~ns at $B=0$~T, see Appendix~\ref{SecApp3}). We observe a small increase in the fidelity for the SiV$^-$ ($\sim 1\%$), and reduction of the gate time, compared to orthogonal scheme at zero magnetic fields. Note that even though the laser intensity is fixed, the gate time varies as the transition dipole overlaps $\langle \psi_i|p_k|\psi_j\rangle$ (which change the effective Rabi frequency and hence the bandwidth) are different for each magnetic field strength.

In the low fidelity range, one or both $\Lambda$-transitions become weakly allowed, while other transitions are driven more strongly. As an example, in Fig.~\ref{fig:5}(b), the regions of longest gate time correspond to weakly allowed $\Lambda$-transitions, which consequently lowers the fidelity in Fig.~\ref{fig:5}(a), for the same magnetic field values. The long gate time of these weakly excited transitions is associated with the transitionless-pulse condition that we explain in Appendix.~\ref{SecApp0}, so for weak effective Rabi frequency, the pulses are narrow-band. In the remaining low fidelity range, one of the $\Lambda$-transitions is weakly driven, thus requiring an increase of the laser power driving the particular transition to match the Rabi frequency of the second $\Lambda$-transition. This is a requirement
that we impose to the CPT transformation to achieve $R_x$ gates. Consequently, with higher laser-power, other bright transitions are driven more strongly, which results in low overall fidelity. Nevertheless, our choice of $\Lambda$-system is not restricted, and for the magnetic field values of low fidelity shown in Fig.~\ref{fig:5}, we could instead select a different $\Lambda$-system.

\section{Control of SnV$^-$ system \label{Sec6}}

\subsection{Zero magnetic fields \label{Sec6a}}

\begin{figure}[!htbp]
    \centering
    \includegraphics[scale=0.55]{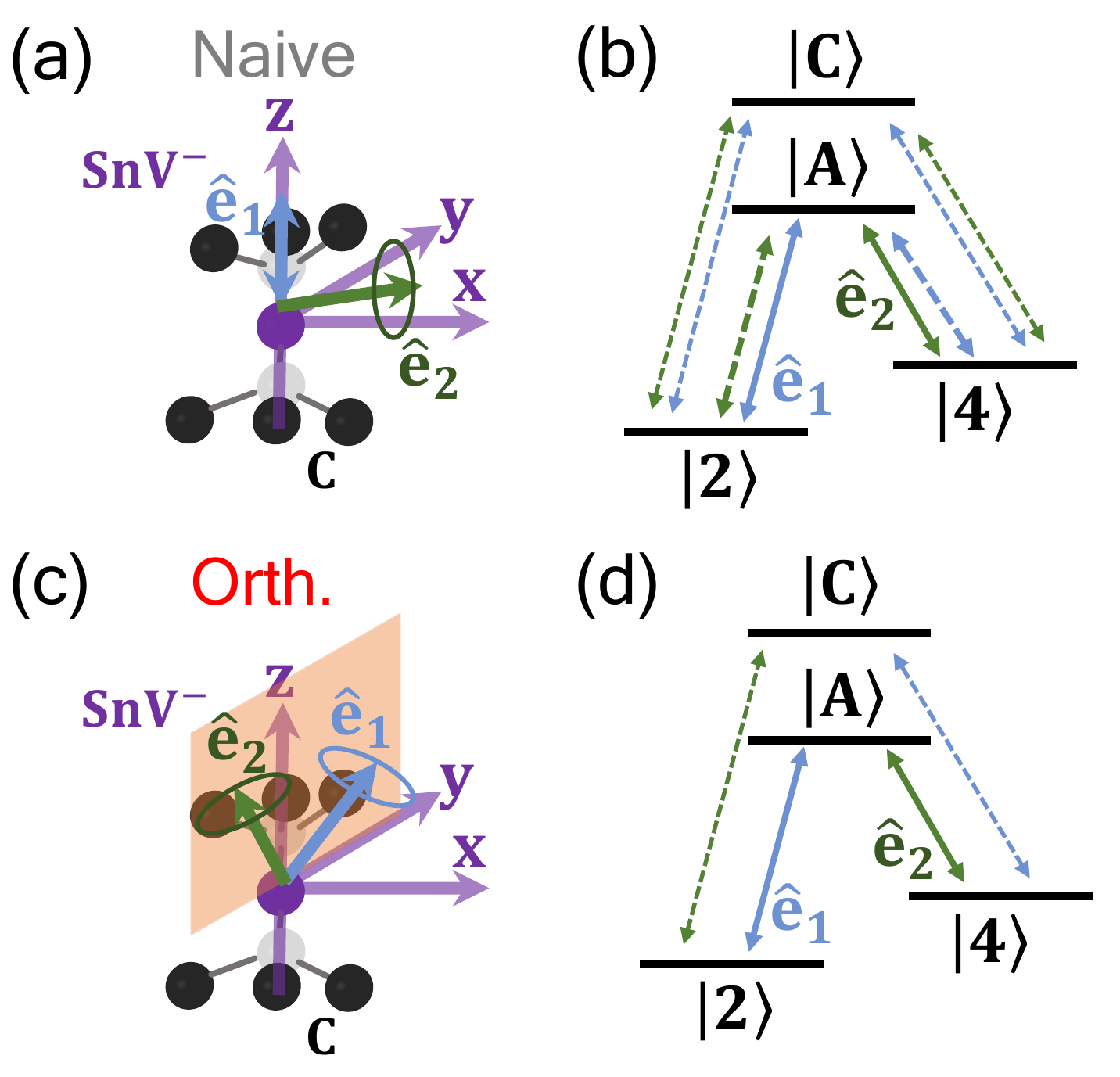}
    \caption{The selection rules-based \textbf{E}-field polarizations of (a) lead to cross-talk and four leakage transitions shown in (b). (c,d) The redefined polarizations in the $yz$-plane can eliminate the cross-talk and additionally two leakage transitions of (b).}
    \label{fig:6}
\end{figure}

The main advantage of the SnV$^-$ defect is its large ground and excited states splittings. This suppresses both incoherent and coherent errors. Due to the large energy separation, orbital relaxations are further suppressed compared to the SiV$^-$, meaning that high fidelity control is possible without millikelvin cooling. The larger splitting also reduces the cross-talk of the lasers driving the transitions. For zero magnetic fields, we find that the bright transitions are A2, C2, B1, D1, B3, D3, A4, and C4. We form the $\Lambda$-system by selecting the $|2\rangle$, $|4\rangle$ and $|\text{A}\rangle$ states, as shown in Fig.~\ref{fig:6}(b). Again, we assume no initialization errors, and in this case, a temperature of $T=6$~K. The spin relaxation time is set to $T_{1,\text{spin}}=1.26$~ms, the orbital relaxation time to $T_{1,\text{orbit}}=38$~ns, the dephasing time to $T_2^*=59$~ns and the optical lifetime to $\tau=4.5$~ns \cite{PhysRevLett.124.023602}. 

\begin{figure*}[!htbp]
    \centering
    \includegraphics[scale=0.52]{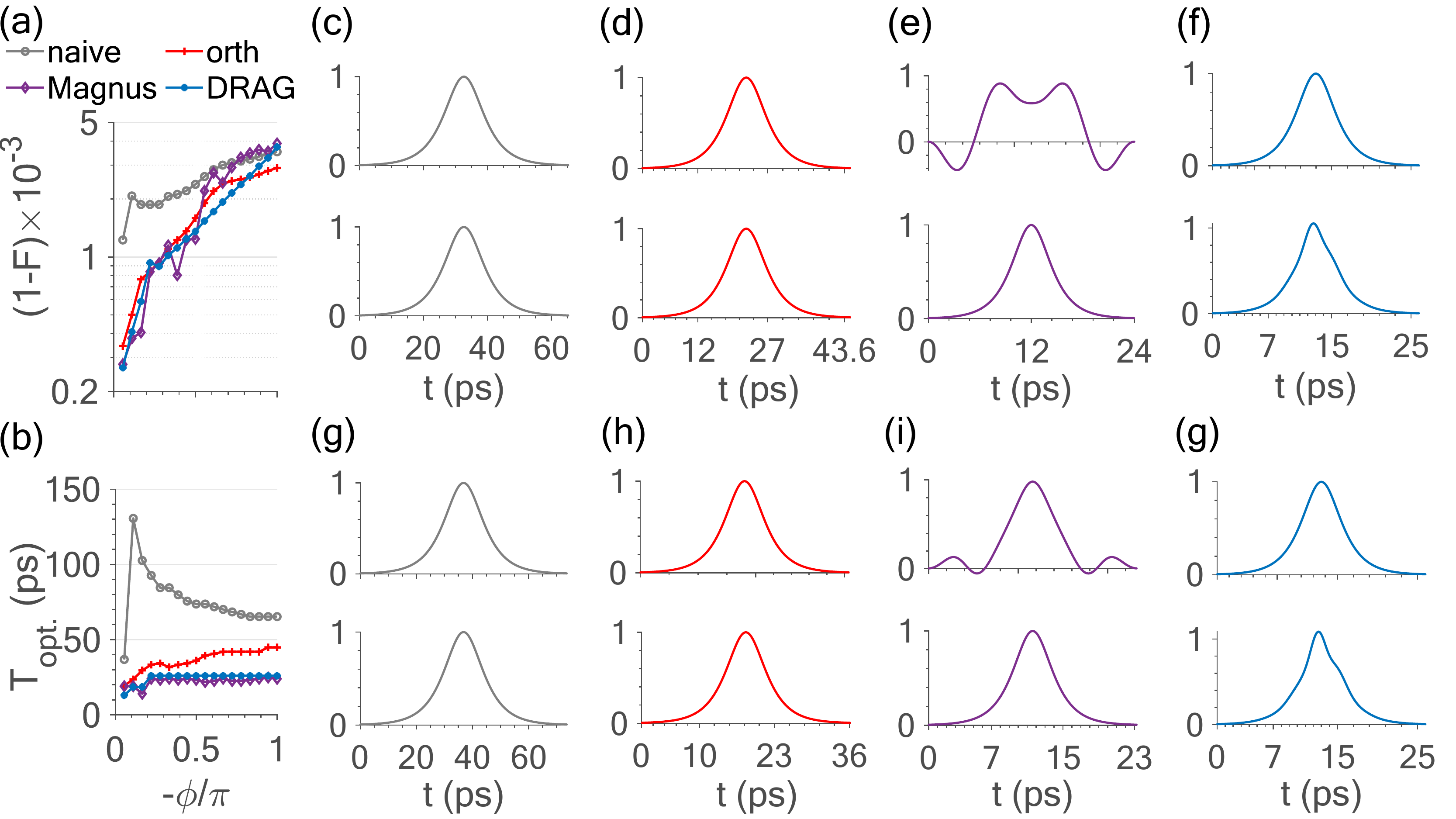}
    \caption{Optical control of SnV$^-$ at $B=0$: Infidelity of $R_x(\phi)$ (a), and optimal gate time (b), corresponding to four different protocols. The pulse envelopes for $R_x(-\pi)$ rotations are shown in (c), (d), (e), (f) and the pulse envelopes for $R_x(-\pi/2)$ are depicted in (g), (h), (i), (g). Gray: naive, red: orthogonal, purple: Magnus and blue: DRAG.  }
    \label{fig:7}
\end{figure*}

In the naive scheme, which is based on selection rules, the A2 transition is strongly driven by $z$-polarized light and the A4 by circularly polarized light [Fig.~\ref{fig:6}(a)]. As a result, both cross-talk and leakage errors are present in the optical control [Fig.~\ref{fig:6}(b)]. However, due to the large ground and excited states splittings of the SnV$^-$, these errors average out more effectively than the SiV$^-$ system, when the pulses are not extremely broadband.

Again, we test the performance of our four protocols; the naive, the orthogonal, the Magnus and the DRAG approaches. All of these schemes can achieve faster and higher fidelity gates compared to the SiV$^-$ system. The large excited state splitting of almost $3$~THz allows to implement the rotations at less than hundreds of picoseconds, while preserving selectivity of the A transitions. Consequently, in contrast to the SiV$^-$, the gates can be performed in a completely relaxation-free duration range, even for the naive approach. Thus, there is no trade-off between frequency selectivity and relaxation errors.

Starting with the naive approach (gray curve) we show the infidelity of $R_x(\phi)$ rotations in Fig.~\ref{fig:7}(a) and the optimal gate time in Fig.~\ref{fig:7}(b). We notice that the gates are well-protected and considerably faster than the SiV$^-$, as the error transitions average out efficiently. The selectivity of the A transitions and the contribution of the cross-talk set a lower bound for the optimal gate time, which on average is greater than 50~ps [see Fig.~\ref{fig:7}(b)].

In the orthogonal scheme, we can remove the cross-talk within the $\Lambda$-system by redefining the polarization of the laser fields. One example would be to select $xz$-polarization for the driving fields (such that $\textbf{E}_1\cdot \textbf{d}_{\text{A4}}=0=\textbf{E}_2\cdot \textbf{d}_{\text{A2}}$), similar to the SiV$^-$ system. However, we found that a different choice of polarizations can additionally eliminate two out of the four leakage transitions. 

We model the Jahn-Teller (JT) contribution according to \cite{PhysRevLett.124.023602}, which gives rise to purely real $\langle p_z\rangle$ and $\langle p_x\rangle$ transition dipoles and purely imaginary $\langle p_y\rangle$ transition dipoles. Under this assumption, and considering $yz$-polarization for $\textbf{E}_1$ and $\textbf{E}_2$, we find that by setting the polarizations to be:

\begin{equation}
    \textbf{E}_1=E_{01}\left(\textbf{y}-\frac{\langle p_y \rangle_{\text{A4}}}{\langle p_z\rangle_{\text{A4}}}\textbf{z}  \right)e^{i(k_1x-\omega_1 t)} +\text{c.c.}
\end{equation}

\begin{equation}
    \textbf{E}_2=E_{02}\left(\textbf{y}-\frac{\langle p_y \rangle_{\text{A2}}}{\langle p_z\rangle_{\text{A2}}}\textbf{z}  \right)e^{i(k_2x-\omega_2 t)} +\text{c.c.},
\end{equation}
we not only resolve the cross-talk, but we also fulfil the relations $\textbf{E}_1 \cdot \textbf{d}_{\text{C2}}=0$ and $\textbf{E}_2\cdot \textbf{d}_{\text{C4}}=0$. Thus, the remaining leakage transitions correspond to the driving of C2 by the $\textbf{E}_2$ field and to the driving of C4 by the $\textbf{E}_1$ field [see Figs.~\ref{fig:6}(c), (d)]. In Appendix~\ref{SecApp2}, we derive the polarizations we need to define to eliminate the cross-talk and two out of the four leakage transitions for arbitrary JT parameters.

With this simple redefinition of the $\textbf{E}$-field polarizations, the orthogonal approach (red curve) achieves enhanced gate fidelities compared to the naive scheme, as shown by the red curve in Fig.~\ref{fig:7}(a). By removing the cross-talk and reducing the leakage, we manage to decrease the optimal gate time below 50~ps [Fig.~\ref{fig:7}(b)]. The optimal pulse envelopes for the $R_x(-\pi)$ and $R_x(-\pi/2)$ are shown in Fig.~\ref{fig:7}(d) and Fig.~\ref{fig:7}(h) respectively, which still correspond to simple sech pulses.

The orthogonal scheme sets the lower bound of gate infidelities and gate durations for unmodulated pulses. To go beyond this limit, we allow for pulse modifications by using the Magnus- and DRAG-based protocols. First, we test the performance of the Magnus protocol. In this method, only the pulse envelope of the $\textbf{E}_1$-field is modified. Even though in the initial frame the control pulse has access only to the C4 error transition and not the C2 (which is driven by the $\textbf{E}_2$-laser field), we find that the linear system we solve to specify the control is still well-defined. More details for the Magnus scheme are given in Appendix~\ref{SecApp5}. The Magnus scheme has the shortest optimal gate-time duration of all methods [see purple curve Fig.~\ref{fig:7}(b)]. Although it seems to underperform compared to the orthogonal scheme for larger rotation angles, the pulses are much more broadband than in the former case. The pulse envelopes are displayed in Fig.~\ref{fig:7}(e) [$R_x(-\pi)$], and Fig.~\ref{fig:7}(i) [$R_x(-\pi/2)$] , where the top panels involve pulse modulation of the laser driving the A2 transition.

Finally, we evaluate the performance of the DRAG protocol. In Fig.~\ref{fig:7}(a), we show with blue curve the infidelity of arbitrary rotations for the DRAG scheme. For almost all rotation angles the infidelity owed to leakage is suppressed, and the gate time is reduced compared to the orthogonal method. The optimal pulse envelopes that implement $R_x(-\pi)$ and $R_x(-\pi/2)$ rotations are shown in Fig.~\ref{fig:7}(f) and Fig.~\ref{fig:7}(g) respectively. In both cases, the modified pulse shown in the bottom panels corresponds to the laser driving the A4 transition. In general, for rotations other than $R_x(-\pi)$, both envelopes require modulation. However, we have performed a simple optimization search on the DRAG corrections, which allows for a redefinition of their amplitude strength. Thus, in this particular case, the optimal solution for the $R_x(-\pi/2)$ gate involves modification of one of the initial driving fields.

\subsection{Non-zero magnetic fields \label{Sec6b}}

Similar to the SiV$^-$, we consider zero two-photon detuning which corresponds to $R_x(\pi)$ rotations, and we select the $\Lambda$-system formed by the states $|1\rangle$, $|3\rangle$ and $|\text{A}\rangle$. For the spin relaxation time, we assume $T_{1,\text{spin}}=150$~ns. We consider the orthogonal approach and in this case we select the $xz$-polarization definition, similar to Sec.~\ref{Sec3}. For non-zero magnetic field strengths, all transitions become bright, and selecting the $yz$-polarizations for the lasers does not offer any advantage (since the transition dipoles are modified due to the Zeeman Hamiltonian). For each magnetic field strength, we would have to select the optimal $\Lambda$-system, and define laser field polarizations that eliminate the cross-talk and also reduce or remove the contribution of the dominant leakage transitions. Although this analysis would be more complete, we instead prefer to showcase the performance of a particular polarization of the $\textbf{E}$-fields of the orthogonal scheme (that mitigates only the cross-talk), and optimize in terms of the magnetic field strengths.

\begin{figure}[!htbp]
    \centering
    \includegraphics[scale=0.45]{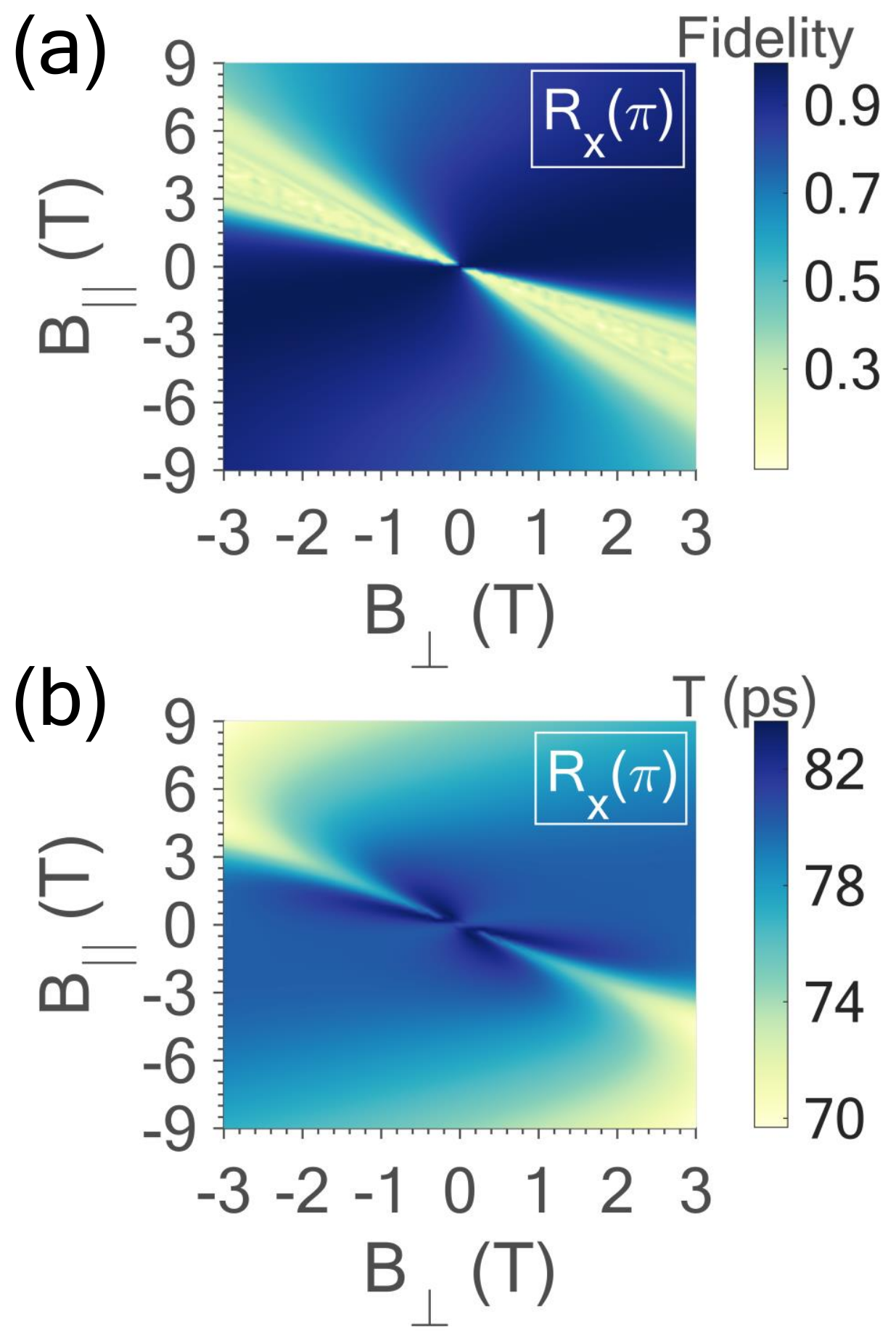}
    \caption{Optical control of SnV$^-$ at $B\neq 0$: Dependence of the fidelity (a) and duration (b) of $R_x(\pi)$ gates on the parallel and perpendicular (with respect to the cryostat axis) magnetic field strengths. For $B\neq0$ we consider the A1-A3 $\Lambda$-system. The regions of low fidelity correspond to weakly excited $\Lambda$-system.}
    \label{fig:8}
\end{figure}

We assume the same magnetic field definitions in the defect coordinate frame as in Sec.~\ref{Sec5b}, and we vary the parallel and perpendicular magnetic field strenghts, with respect to the cryostat axis. In Fig.~\ref{fig:8}(a) we show the fidelity and in Fig.~\ref{fig:8}(b) the gate time of $R_x(\pi)$ rotations for the optimal laser field intensity. The maximum fidelity corresponds to $F_{\text{max}}^{\text{SnV$^-$}}=0.996$ for $B_\perp=0.3$~T and $B_\parallel=0.2$~T and the gate time is $T=80$~ps. This is also the maximum achievable fidelity for the $xz$-polarization in the absence of magnetic field strengths, and the gate time is also close to the zero magnetic field case [see Appendix~\ref{SecApp3}].

The low fidelity range arises for the same reasons as in the case of the SiV$^-$ system. Specifically, the $\Lambda$-transitions are weakly excited and our choice of $\Lambda$-system is not optimal for the particular magnetic fields. As we mentioned for the SiV$^-$ system, even though the laser field intensity is fixed in this case, the bandwidth of the pulse varies, since the transition dipoles depend on the Zeeman Hamiltonian term. For the magnetic field strengths where the fidelity is low, the choice of a different $\Lambda$-system could still maintain high fidelity control.

\section{Conclusions}

In conclusion, we have designed optical control protocols for high-fidelity rotations of two defect systems: the SiV$^-$ and SnV$^-$ in diamond. We use coherent population trapping techniques combined with judicious choice of laser polarizations to mitigate the cross-talk issue of the $\Lambda$-transitions caused by the Jahn-Teller effect. Importantly, strain induced due to integration of the defects in photonic structures can result in enhanced orbital mixing and modification of the selection rules, and hence, more intensified cross-talk. Thus, our cross-talk elimination approach could also be beneficial in such a context. We implement simulations of arbitrary rotations both in the absence and presence of external magnetic fields and thoroughly test the maximum fidelity that we can reach without any additional corrections. To an extended generalization, the choice of polarizations can ensure both vanishing cross-talk and reduced number of leakage transitions.

For the SiV$^-$, there is a trade-off between faster gates protected from relaxation and slower gates protected from leakage errors. On the contrary, for the SnV$^-$, we can safely reach the gate time range where the dissipation mechanisms are  negligible, without causing enhanced leakage.
We show that with our orthogonal pulse scheme we achieve fast and high fidelity control for the SnV$^-$ system, due to its larger ground and excited state splittings. 

Further, we use a Magnus expansion technique, as well as a newly developed version of the DRAG technique, to mitigate leakage errors when considering broadband pulses. The corrective modifications in the Magnus and DRAG schemes involve simple cosine envelopes that can be generated using arbitrary waveform generators and electro-optical modulators, which create modulated pulses from a CW laser. In general, pulses carved out of a CW laser have limited power and speed, but a power enhancement could be achieved with a fast response optical amplifier (e.g. semiconductor amplifiers with up to tens of GHZ repetition rate \cite{10.1117/12.2256791}). Depending on experimental constraints (e.g. laser power, duration), one could select the least demanding and most practical approach to counteract leakage errors.

\begin{acknowledgments}
The authors would like to thank Shahriar Aghaei, Jonas Becker, Alison Emiko Rugar, Jelena Vuckovic, as well as Arian Vevzaee  for valuable discussions. The authors were supported by the United States National Science Foundation under the grant 1838976.

\end{acknowledgments}

\appendix

\section{CPT control with sech pulses \label{SecApp0}}

As we mentioned in the main text, the destructive interference in the CPT scheme leads to a dark state that is completely decoupled from the dynamics of the three-level system. This is achieved by tuning the laser parameters (relative amplitudes and phases), and satisfying the two-photon resonance condition $(\Delta_1=\Delta_2\equiv\Delta)$, where $\Delta_\text{j}$ is the detuning of the transition labeled $j$. The mapping from the initial qubit states in the lab frame to the dark-bright basis can be performed via the transformation:
\begin{equation}\label{Eq1}
    R_{\text{db}}=\begin{pmatrix}
    \cos \frac{\theta}{2} & -e^{-i\alpha}\sin\frac{\theta}{2} \\
    e^{i\alpha}\sin\frac{\theta}{2} & \cos\frac{\theta}{2}
    \end{pmatrix}.
\end{equation}
Effectively, this transformation  defines the rotation axis of the qubit, which is $\textbf{n}=(\sin\theta\cos  \alpha,\sin\theta \sin\alpha,\cos\theta)$, while it also enables the reduction of the initial problem into a two-level system. In particular, the Hamiltonian in the dark-bright basis reads:
\begin{equation}\label{Eq2}
    H_{\text{db}}=\Omega_{\text{eff}}f(t)e^{i\Delta t}\sigma_{\text{be}}+\text{H.c.},
\end{equation}
where $\sigma_{\text{be}}=|\text{b}\rangle \langle \text{e}|$, with $|\text{b}\rangle$ being the bright state and $|\text{e}\rangle$ the excited state. Also, the effective Rabi frequency in this frame is expressed in terms of the original Rabi frequencies as $\Omega_{\text{eff}}=\sqrt{|\Omega_1|^2+|\Omega_2|^2}$. For a general pulse envelope, the two-level problem is not analytically solvable. Here we consider hyperbolic secant pulses (i.e. $f(t)=\text{sech}(\sigma t)$), that have been proven  to be  analytically solvable \cite{PhysRev.40.502}, and lead to a rotation in the qubit subspace given by \cite{PhysRevB.74.205415,EconomouPRL2007}:
\begin{equation}\label{Eq3}
    U_0=\begin{pmatrix}
    1 & 0 \\
    0 & e^{-i\phi}
    \end{pmatrix}.
\end{equation}

As shown by Eq.~(\ref{Eq3}), the dark state does not evolve, whereas the bright state picks up a phase given by $\phi=2\tan^{-1}(\sigma/\Delta)$, where $\Delta$ is the detuning and $\sigma$ the bandwidth. Control of both rotation axis and angle is achieved by combining CPT with hyperbolic secant pulses, which allows us to design arbitrary single-qubit gates. The only additional requirement is that the bandwidth is equal to the effective Rabi frequency in the CPT frame ($\sigma=\Omega_{\text{eff}}$), such that the pulse is transitionless, i.e. the population returns to the ground states at the end of the pulse.

\renewcommand\thefigure{\thesection.\arabic{figure}}    

\setcounter{figure}{0}    

\section{Details of the simulations \label{SecApp1}}

\begin{figure}[!htbp]
    \centering
    \includegraphics[scale=0.35]{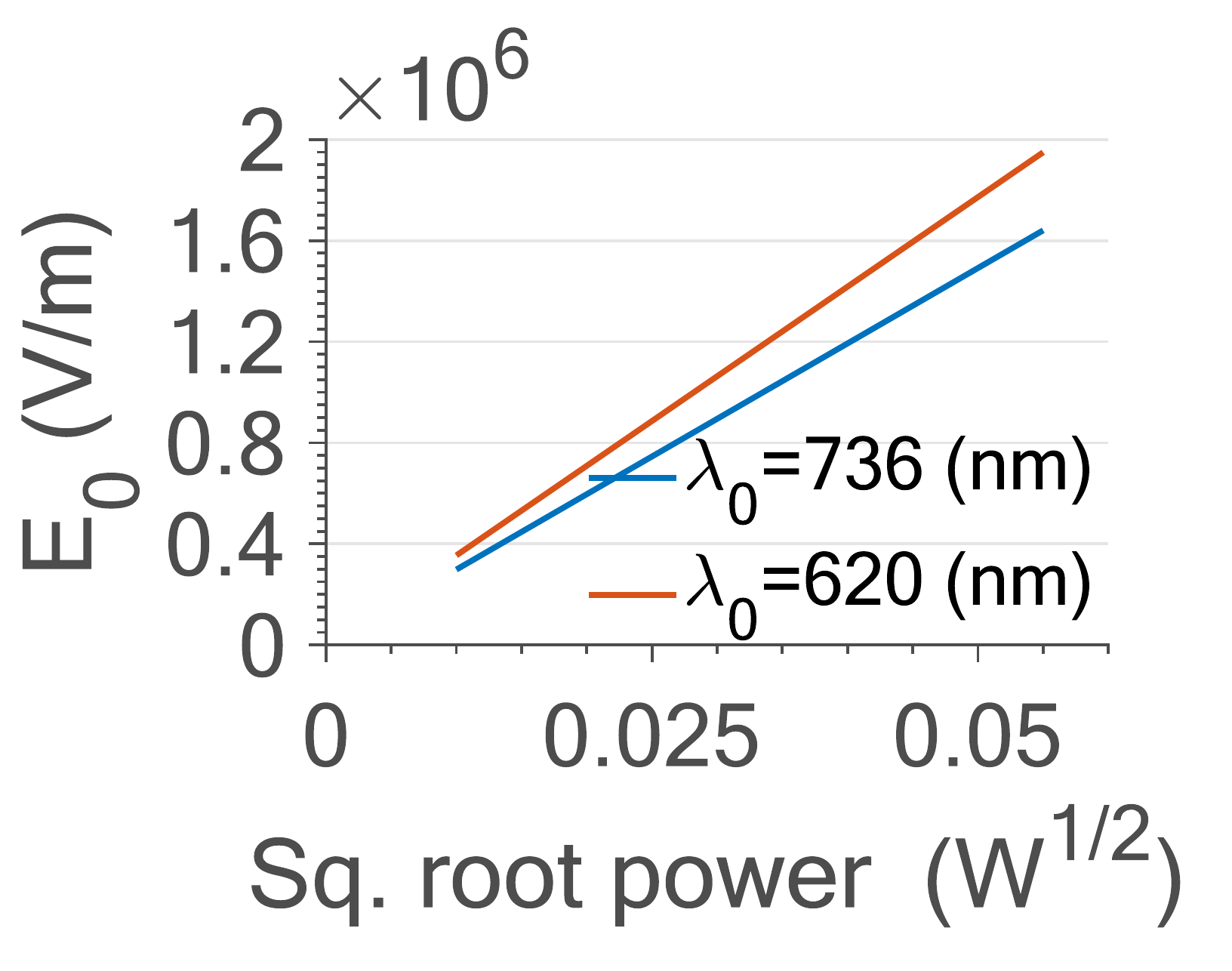}
    \caption{Electric field amplitude versus the square root of the laser power, for the SiV$^-$ central wavelength (blue) and the SnV$^-$ central wavelength (red).}
    \label{fig:Ap1}
\end{figure}

In this section, we provide further details regarding our simulations. First, based on Ref.~\cite{Zhang:17}, we use the laser power applied on the SiV$^-$ system for $\pi$-rotations to calculate the electric field amplitude and estimate the Rabi frequencies. For a numerical aperture NA=0.7, the spot-size (radius) of the laser is given by $w_0=\lambda_0/(\pi \text{NA})$, where $\lambda_0$ is the wavelength of a specific transition for the SiV$^-$ or the SnV$^-$, which can be assumed to be close to the central transition. For the SiV$^-$ the central wavelength is $\lambda_0\approx736$~nm, while for the SnV$^-$ it is $\lambda_0\approx 620$~nm. Assuming an emitter  focused at the center of the beam, the intensity is related with the power and spot-size by the expression:
\begin{equation}\label{EqAp1}
    I=\frac{P_0}{\pi w_0^2},
\end{equation}
while it can also be expressed as:
\begin{equation}\label{EqAp2}
    I=\frac{cn\epsilon_0}{2}|E_0|^2,
\end{equation}
where $c$ is the speed of light, $n=2.4$ is the refractive of diamond, $E_0$ is the electric field amplitude and $\epsilon_0$ is the vacuum permittivity. The factor of 1/2 comes from averaging the intensity. By combining Eq.~(\ref{EqAp1}) with Eq.~(\ref{EqAp2}) we can express the electric field in terms of the laser power as:
\begin{equation}\label{EqAp3}
    |E_0|=\sqrt{\frac{2P_0}{\pi w_0^2 c n \epsilon_0}}.
\end{equation}
From Eq.~(\ref{EqAp3}), we calculate the electric field amplitude based on the laser powers of Ref.~\cite{Zhang:17}, shown in Fig.~\ref{fig:Ap1}. For the SiV$^-$, the maximum electric field amplitude we have considered is $E_0=8.5\times 10^4$ (V/m), while for the SnV$^-$, we have considered up to $E_0\approx 1\times 10^6$ (V/m). (These values exclude the numerically optimized DRAG pulses, whose amplitude corrections have a multiplicative factor $|c|<4$). Further, for the $z$-transition dipoles, we have taken into account the experimental enhancement factor of 2 of the $z$-dipole. In general, the optimal ranges of operation we found for both defects are smaller than these maximum $E_0$ amplitudes, so the laser power should correspond to the experimentally safe and achievable ranges.

Further, we calculate the Rabi frequencies for each transition as:
\begin{equation}
    \Omega^{ij}=\alpha\frac{e r_0 |E_0|}{ \hbar}\langle \psi_i |p_k|\psi_j\rangle,
\end{equation}
where $e$ is the electronic charge, and for $r_0$ we assume $r_0=0.53$~\text{\AA}. We estimate the multiplicative factor $\alpha\approx 6.663$ for the SiV$^-$ and $\alpha\approx3.3$ for the SnV$^-$. In particular, for the SiV$^-$, that would give rise to a dipole moment of approximately $\mu\approx 16.6$~Debye, which is close to $\mu=14.3$~Debye reported in \cite{https://doi.org/10.1002/pssa.201700586}. Also, $\langle \psi_i | p_k|\psi_j\rangle$, is the dipole overlap of the transition and the matrices $p_x$, $p_y$, $p_z$ are given by group theory \cite{PhysRevLett.112.036405}. We should mention that in our simulations we start by defining the driving Hamiltonian without the factor of $1/2$ in front of the Rabi-frequencies, which would be a result of the RWA. This means that the $E_0$ value should be twice as much as the approximate values we report above. To calculate the eigensystem of all eight levels of each defect, we consider three main interaction terms: the spin-orbit coupling, the Jahn-Teller and the Zeeman effects.

Regarding the Lindblad relaxation operators, we follow a similar convention as in Ref.~\cite{Mete2017}. First, for the dephasing mechanism, we assume an equal dephasing of all states:
\begin{equation}
    G_{\text{deph}}=\frac{1}{\sqrt{T_2^*}}|i \rangle\langle i|.
\end{equation}
Spin-relaxation mechanisms lead to a change of the spin-state while preserving the same orbital part, which occur within the ground or excited state manifold. We define the Lindblad spin-relaxation operators as:
\begin{equation}
    G_{\text{spin}}=\frac{1}{\sqrt{2T_{1,\text{spin}}}}|i\rangle\langle j|.
\end{equation}
Orbital dissipation mechanisms occur between different orbital states of the same spin projection. We define the associated Lindblad operator as:
\begin{equation}
    G_{\text{orbit}}=\sqrt{F_{\text{orbit}}}|i\rangle\langle j|,
\end{equation}
where the decay rate $F_{\text{orbit}}$ is different for an upward or downward relaxation. For a downward relaxation we define:
\begin{equation}
    F_{\text{orbit, down}}=\frac{1}{T_{1,\text{orbit}}(1+e^{-|\Delta E|/k_b T})}
\end{equation}
and for an upward:
\begin{equation}
    F_{\text{orbit,up}}=F_{\text{orbit,down}}e^{-|\Delta E|/k_bT},
\end{equation}
i.e. the orbital relaxations are scaled by Boltzman factors, with the upward orbital relaxations being less probable. $\Delta E$ is the energy difference between the levels that participate in the orbital relaxation mechanism. Finally, for the lifetime relaxations, we define the Lindblad operators as:
\begin{equation}
    G_{\text{lifetime}}=\frac{1}{\sqrt{\tau}}|i\rangle\langle j|,
\end{equation}
where $\sigma_{ij}=|i\rangle\langle j|$ corresponds to a bright transition, and $\tau$ is the lifetime.

\setcounter{figure}{0}

\section{General method for removing the cross-talk and one leakage transition \label{SecApp2}}

As we mentioned in the main text, we can always redefine the polarization of the laser fields to remove completely the cross-talk within the $\Lambda$-system. We also mentioned that for the SnV$^-$ and by using the Jahn-Teller (JT) parameters of \cite{PhysRevLett.124.023602}, the polarization of the \textbf{E}-fields was found to be additionally orthogonal to one leakage transition each.

However, this extra property depends on the modeling of the JT interaction. To resolve this subtlety, we derive analytically the transition dipoles for arbitrary JT parameters. Assuming no crystal strain, and working at $B=0$~T, the only non-zero interaction terms are the spin-orbit coupling and the JT effect. 

By expressing the interaction Hamiltonian in the $|e_\pm\rangle$ orbital basis and in the $\{|\uparrow\rangle, |\downarrow\rangle\}$ spin basis, the two interaction terms read:

\begin{equation}
    H_{\text{g/e}}=\begin{pmatrix}
    Q_{x,\text{g/e}} & Q_{y,\text{g/e}} \\
    Q_{y,\text{g/e}} & -Q_{x,\text{g/e}}
    \end{pmatrix}\otimes \textbf{1}-\frac{\lambda_{\text{SO},\text{g/e}}}{2}L_z\otimes S_z,
\end{equation}
where $Q_{x,\text{g/e}}=Q_{\text{g/e}} \cos \phi_{\text{g/e}}$, $Q_{y,\text{g/e}}=Q_{\text{g/e}}\sin \phi_{\text{g/e}}$, $\lambda_{\text{SO}_{\text{g/e}}}=\Delta E_{\text{g/e}} \cos \theta_{\text{g/e}}$ and $Q_{\text{g/e}}=\Delta E_{\text{g/e}}/2 \sin \theta_{\text{g/e}}$. The subscript $g$ and $e$ denote ground and excited state respectively. The parameter $\theta_{\text{g/e}}$ can be tuned so as to give the relative strength of the SO and JT contributions that are experimentally observed. The unnormalized eigenvectors are given by:
\begin{eqnarray}
    v_{1,\text{g/e}}&=&\begin{pmatrix}
    0 &
    e^{i\phi_{\text{g/e}}}\tan \frac{\theta_{\text{g/e}}}{2} & 0 & 1\end{pmatrix}^T \\
    v_{2,\text{g/e}}&=&\begin{pmatrix}
    e^{i\phi_{\text{g/e}}}\cot \frac{\theta_{\text{g/e}}}{2} &
    0 & 1 & 0\end{pmatrix}^T \\
    v_{3,\text{g/e}}&=&\begin{pmatrix}
    0 &
    -e^{i\phi_{\text{g/e}}}\cot \frac{\theta_{\text{g/e}}}{2} & 0 & 1\end{pmatrix}^T \\
    v_{4,\text{g/e}}&=&\begin{pmatrix}
    -e^{i\phi_{\text{g/e}}}\cot \frac{\theta_{\text{g/e}}}{2} &
    0 & 1 & 0\end{pmatrix}^T,
\end{eqnarray}
where $v_{1,\text{g/e}}$ and $v_{2,\text{g/e}}$ correspond to the eigenenergy $-\Delta E_{\text{g/e}}/2$ and the eigenvectors $v_{3,\text{g/e}}$ and $v_{4,\text{g/e}}$ to the eigenenergy $\Delta E_{\text{g/e}}/2$. 

As an example let's assume that we use the $\Lambda$-transitions A2 and A4, and that we want to make the $\textbf{E}_1$ field orthogonal to $\textbf{d}_{\text{A4}}$ and $\textbf{d}_{\text{C2}}$. Under this notation, the non-zero transitions would be between $v_{1,\text{g}} \leftrightarrow v_{1,\text{e}}$ and $v_{1,\text{g}}\leftrightarrow v_{3,\text{e}}$. Thus, the transition dipole $\textbf{d}_1=\textbf{d}_{v_{1,\text{g}}v_{1,\text{e}}}$ (of A2) is:

\begin{equation}
    \textbf{d}_1=\frac{1}{|\sec \frac{\theta_\text{e}}{2} \sec \frac{\theta_\text{g}}{2}|}\begin{pmatrix}
    -(e^{-i\phi_\text{e}} \tan \frac{\theta_\text{e}}{2} + e^{i\phi_\text{g}} \tan \frac{\theta_\text{g}}{2})  \\
    i(e^{i\phi_\text{e}}\tan\frac{\theta_\text{e}}{2} -e^{i\phi_\text{g}} \tan\frac{\theta_\text{g}}{2} ) \\
    2(1+ e^{i(\phi_\text{e}+\phi_\text{g})} \tan \frac{\theta_\text{e}}{2}\tan\frac{\theta_\text{g}}{2}  )
    \end{pmatrix}.
\end{equation}
Similarly the dipole $\textbf{d}_2=\textbf{d}_{v_{1,\text{g}}v_{3,\text{e}}}$ (of C2) is:

\begin{equation}
    \textbf{d}_2=\frac{1}{|\sec \frac{\theta_\text{e}}{2} \sec \frac{\theta_\text{g}}{2}|}\begin{pmatrix}
    (e^{i\phi_{\text{e}}} \cot \frac{\theta_\text{e}}{2} - e^{i\phi_\text{g}} \tan \frac{\theta_\text{g}}{2})  \\
    -i(e^{i\phi_\text{e}}\cot\frac{\theta_\text{e}}{2} +e^{i\phi_\text{g}} \tan\frac{\theta_\text{g}}{2} ) \\
    2(-1+ e^{i(\phi_\text{e}+\phi_\text{g})} \cot \frac{\theta_\text{e}}{2}\tan\frac{\theta_\text{g}}{2}  )
    \end{pmatrix}.
\end{equation}
Finally, the dipole $\textbf{d}_3=\textbf{d}_{v_{3,\text{g}}v_{1,\text{e}}}$ (of A4) is:

\begin{equation}
     \textbf{d}_3=\frac{1}{|\sec \frac{\theta_\text{e}}{2} \sec \frac{\theta_\text{g}}{2}|}\begin{pmatrix}
    e^{i\phi_{\text{g}}}(\cot\theta_\text{g} +\csc\theta_\text{g})-e^{i\phi_{\text{e}}}\tan\frac{\theta_{\text{e}}}{2}  \\
    -i( e^{i\phi_{\text{g}}}(\cot\theta_{\text{g}}+\csc\theta_{\text{g}})+e^{i\phi_{\text{e}}}\tan\frac{\theta_{\text{e}}}{2}  ) \\
    2(1- e^{i(\phi_{\text{e}}+\phi_{\text{g}})} \cot \frac{\theta_{\text{g}}}{2}\tan\frac{\theta_{\text{e}}}{2}  )
    \end{pmatrix}.
\end{equation}

The goal is to make $\textbf{E}_1\cdot \textbf{d}_2=0$ and $\textbf{E}_1\cdot \textbf{d}_3=0$.  Thus, considering the general expression:

\begin{equation}
    \textbf{E}_1=E_{01}(c_1 \textbf{x}+c_2\textbf{y}+c_3\textbf{z})e^{i(\textbf{k}\cdot \textbf{r}-\omega t)}+\text{c.c.},
\end{equation}
we specify the $c_2$ and $c_3$ that satisfy the orthogonality relations:
\begin{equation}
    c_3=\frac{-ic_1(\cos\theta_{\text{e}}+\cos\theta_{\text{g}})}{2(\sin\theta_{\text{e}}\sin\phi_{\text{e}}+\sin\theta_{\text{g}}\sin\phi_{\text{g}})}
\end{equation}
\begin{equation}
    c_2=\frac{c_1(-\cos\phi_{\text{e}}\sin\theta_{\text{e}} + \cos\phi_{\text{g}}\sin\theta_{\text{g}})}{\sin\theta_{\text{e}}\sin\phi_{\text{e}}+\sin\theta_{\text{g}}\sin\phi_{\text{g}}}.
\end{equation}
Similarly, we could follow the same procedure to satisfy $\textbf{E}_2\cdot \textbf{d}_{\text{A2}}=0=\textbf{E}_2\cdot \textbf{d}_{\text{C4}}$. 

Alternatively, we could choose the polarization of $\textbf{E}_1$ such that we satisfy the orthogonality relation to the A4 $\Lambda$-transition while also minimizing both leakage transitions C2 and C4 that are driven by each laser field (and similarly for the $\textbf{E}_2$ field).

\section{Gate time dependence of the fidelity \label{SecApp3}}

\begin{figure}[!htbp]
    \centering
    \includegraphics[scale=0.37]{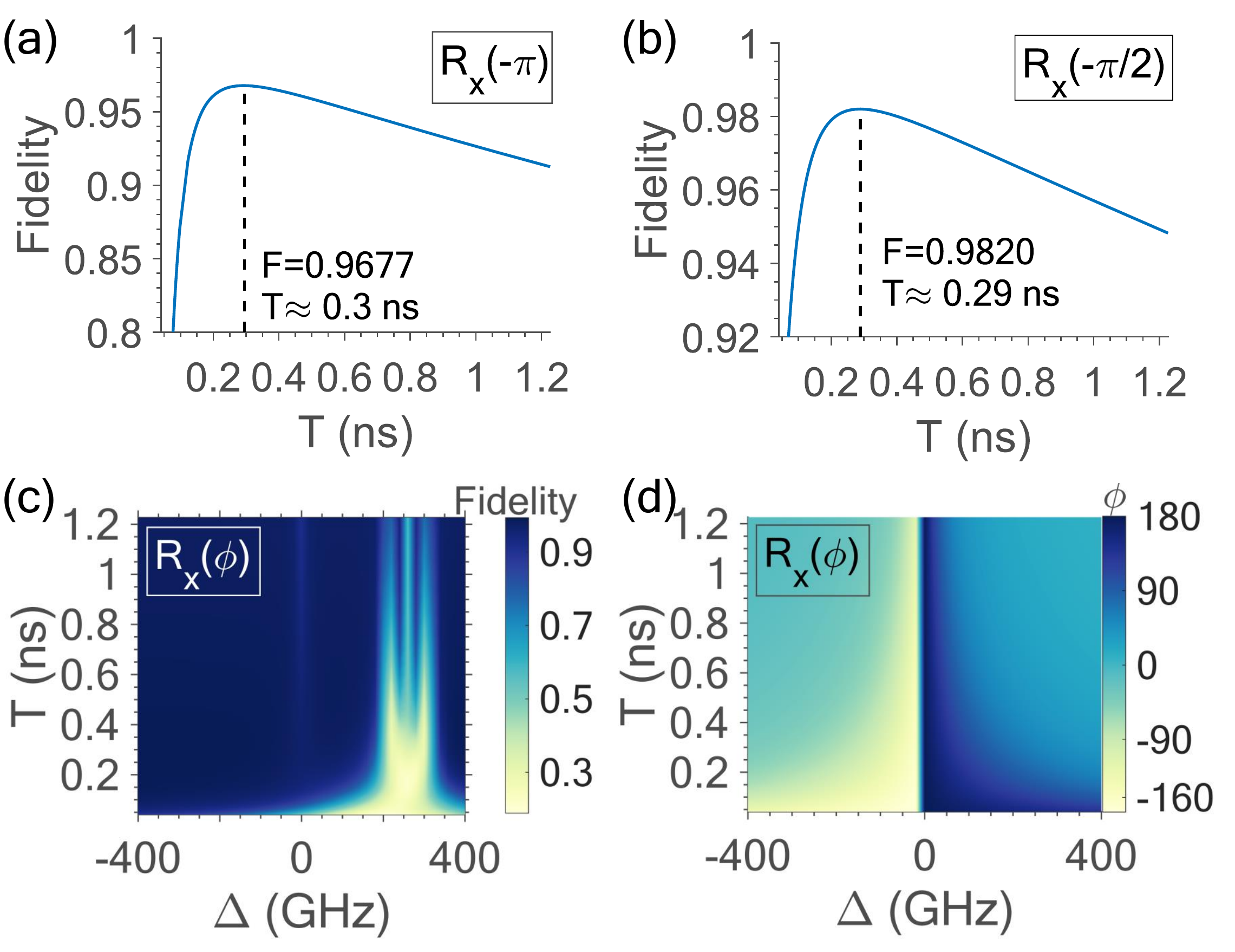}
    \caption{Gate time dependence of the fidelity for $R_x(-\pi)$ (a) and $R_x(-\pi/2)$ gates for the SiV$^-$. Fidelity of $R_x(\phi)$ rotations (c) and rotation angles (d) versus the gate time and two-photon detuning.}
    \label{App:SiV}
\end{figure}

Here we show the time dependence of the fidelity of the gates for the SiV$^-$ and the SnV$^-$ defects. For both systems we consider the orthogonal scheme and use a combination of $xz$-polarizations for the \textbf{E}-fields (such that we cancel the cross-talk errors). In Fig.~\ref{App:SiV} (a) and Fig.~\ref{App:SiV}(b) we show the fidelity versus the gate time for $R_x(\pi)$ and $R_x(-\pi/2)$ rotations for the SiV$^-$.
Narrowband pulses suffer from relaxation errors, while significantly broadband pulses suffer from enhanced leakage errors. The optimal gate time for both rotations is $T=0.3$~ns, with a fidelity close to $F=0.97-0.98$. In Fig.~\ref{App:SiV}(c) we show the fidelity of arbitrary rotations versus the gate time and two-photon detuning. For $\Delta\gtrsim 200$~GHz, which corresponds to the excited state splitting, the upper-excited manifold is driven more strongly leading to significant leakage errors. Nevertheless, the same rotation angles can be implemented with negative detuning at high fidelities. The corresponding rotation angles are shown in Fig.~\ref{App:SiV}(d).

\begin{figure}[!htbp]
    \centering
    \includegraphics[scale=0.37]{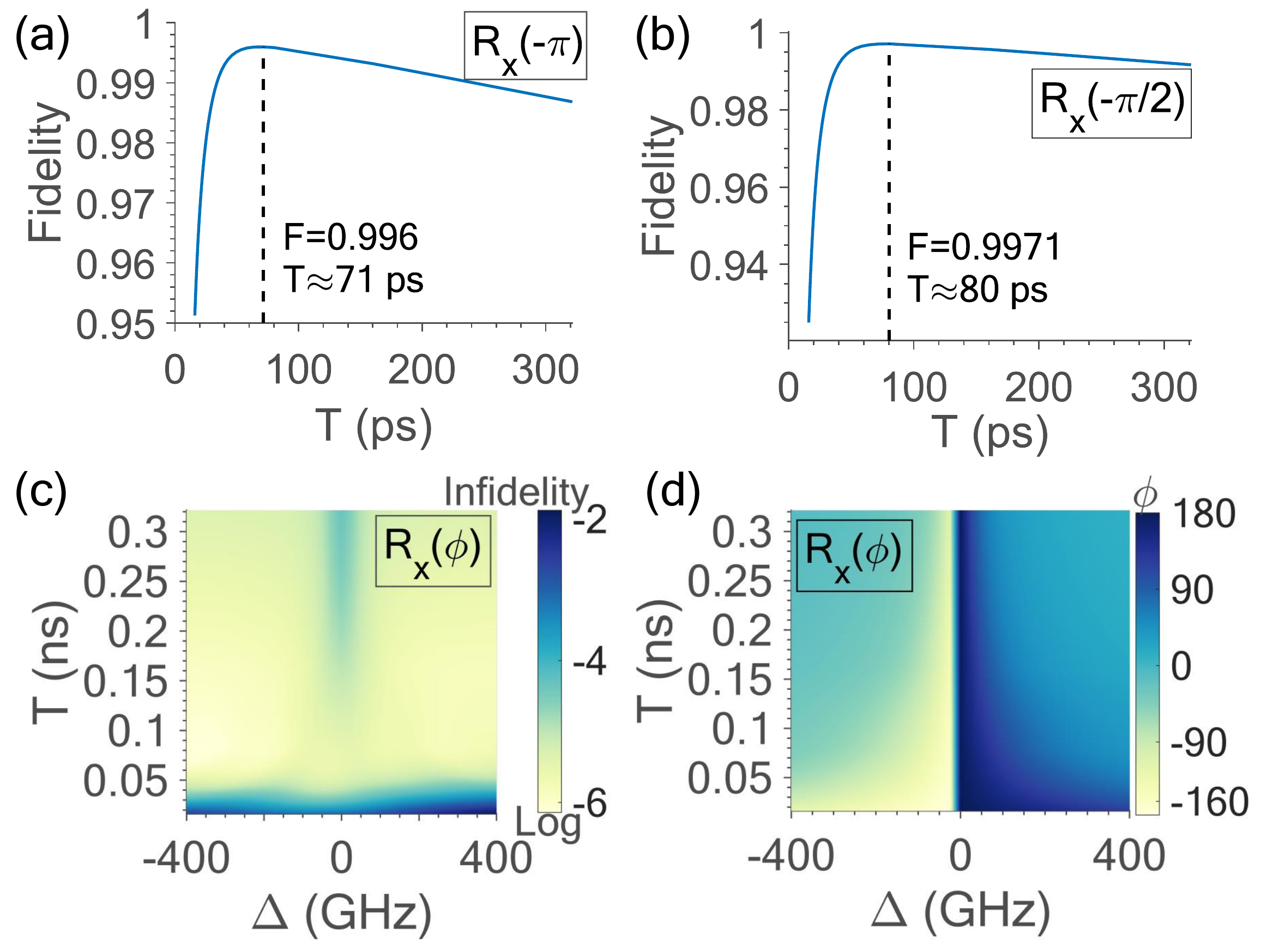}
    \caption{Gate time dependence of the fidelity for $R_x(-\pi)$ (a) and $R_x(-\pi/2)$ gates for the SnV$^-$. Fidelity of $R_x(\phi)$ rotations (c) and rotation angles (d) versus the gate time and two-photon detuning.}
    \label{App:SnV}
\end{figure}

Similarly, we show the gate time dependence of $R_x(\pi)$ and $R_x(-\pi/2)$ for the SnV$^-$ system in Fig.~\ref{App:SnV}(a) and Fig.~\ref{App:SnV}(b).
The infidelity of arbitrary rotations in logarithmic scale is shown in Fig.~\ref{App:SnV}(c) and the rotation angles are shown in Fig.~\ref{App:SnV}(d). In this case, due to the large excited state splitting of the defect, the positive rotation angles exhibit low infidelity.

\setcounter{figure}{0}    
\section{Effect of relaxations on the fidelity  \label{SecApp4}}

\begin{figure}[!htbp]
    \centering
    \includegraphics[scale=0.38]{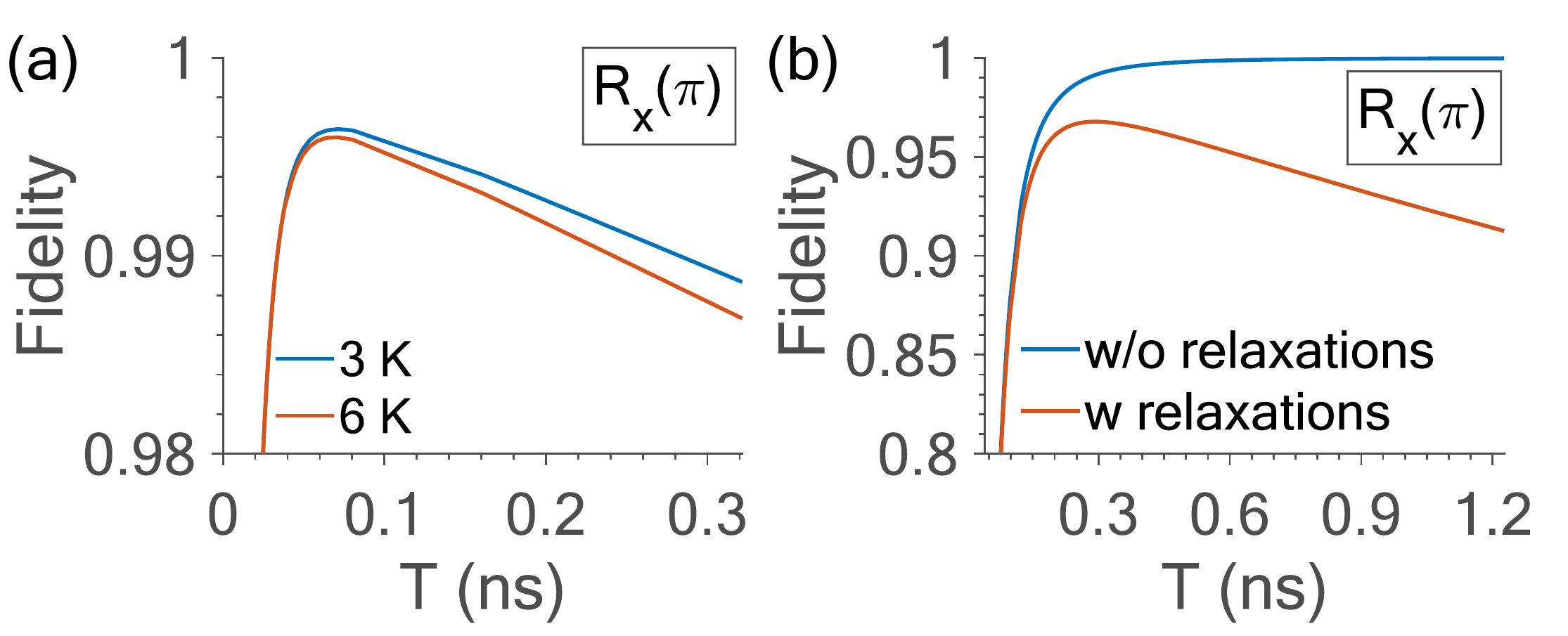}
    \caption{(a) Fidelity of $R_x(\pi)$ rotations at zero magnetic fields for the SnV$^-$, for temperature $T=3$~K (blue) and $T=6$~K (red). For a gate time $T<$0.1~ns there is no temperature dependence of the fidelity, as below this gate time the relaxations seize to contribute.  (b) Fidelity of $R_x(\pi)$ rotations for the SiV$^-$ in the presence of relaxations (red) and without dissipation mechanisms (blue).}
    \label{fig:App3}
\end{figure}

In Fig.~\ref{fig:App3}(a), we test the fidelity of $R_x(\pi)$ rotations (orthogonal scheme) for the case of the SnV$^-$, considering two different temperatures. For $T=3$~K, we assume $T_2^*=540$~ns and $T_{1,\text{spin}}=10.2$~ms, while for $T=6$~K we assume $T_{1,\text{spin}}=1.26$~ms and $T_2^*=59$~ns. We observe that the two curves are almost identical for gate times $T< 0.1$~ns, while for longer gates the deviation starts to increase further. 

In general, for all rotation angles, the optimal fidelity range should lie below $\sim 0.1$~ns for the SnV$^-$, such that the contribution of the relaxations is negligible and the fidelity is almost independent of the temperature. On the other hand, for the SiV$^-$, the optimal range is shifted to longer times, as the leakage errors tend to increase substantially for broadband pulses [Fig.~\ref{fig:App3}(b)]. (Again we show the performance of the orthogonal scheme.) Upon cooling to mK temperatures, the phonon induced relaxations can be suppressed substantially \cite{PhysRevLett.119.223602}, since the qubit states become decoupled from the phonon bath.

\section{Corrections to leakage errors with the Magnus expansion approach \label{SecApp5}}

In this section, we provide the linear system of equations of the Magnus methods, for both defect systems. According to the Magnus expansion approach of \cite{Clerk2021}, we specify the control fields by reducing the problem to a linear set of equations.  We consider zero external magnetic fields since, in this case, we have a smaller number of unwanted transitions that we want to cancel out. We further distinguish two cases: i) resonant driving ($R_x(\pi)$ rotations) and ii) off-resonant driving ($R_x(\phi)$ rotations). As we will show in subsequence, we can generalize from the resonant to the off-resonant case by a slight modification of our linear system of equations.

Starting from the SiV$^{-}$ system, our goal is to find a corrective Hamiltonian $W(t)$ to suppress the leakage errors. In the main text, we chose the $\Lambda$-system formed by the states $|1\rangle , |4\rangle $ and $|\text{A}\rangle$. Assuming perfect initialization, the main error of our orthogonal scheme is the driving of the C1 and C4 transitions, which leads to leakage outside of our $\Lambda$-system. 

We first decompose the error terms of our Hamiltonian into the Gell-Mann basis. Starting from the lab frame, we fix two orthogonal polarizations for the $\Lambda$-transitions, as described in the main text. Thus, in the interaction frame, our initial Hamiltonian including only the A1, A4 and C1 and C4 transitions reads:
\begin{widetext}
\begin{equation}
\begin{split}
    H&=(|\Omega_{1}^{\text{A1}}|e^{i\phi_1}e^{i\Delta t}\sigma_{1\text{A}}+
    \Omega_{1}^{\text{C1}}e^{i(\Delta-\delta_{\text{es}})t}\sigma_{1\text{C}}+
    |\Omega_{1}^{\text{C4}}|e^{i\phi_{\text{C}4}}e^{i(\Delta-(\delta_{\text{es}}-\delta_{\text{gs}}))t}\sigma_{4\text{C}}+
    \text{H.c.})\\&+
    (|\Omega_{2}^{\text{A4}}|e^{i\phi_1}e^{i\Delta t}\sigma_{4\text{A}}+
    |\Omega_2^{\text{C1}}|e^{-i\phi_{\text{C}4}}e^{i(\Delta-(\delta_{\text{es}}+\delta_{\text{gs}}))t}\sigma_{1\text{C}}-
    |\Omega_{2}^{\text{C4}}|e^{i(\Delta-\delta_{\text{es}})t}\sigma_{4\text{C}} +
    \text{H.c.})f(t),
    \end{split}
\end{equation}
\end{widetext}
where $\sigma_{ij}=|i\rangle\langle j|$, $\delta_{\text{es}}=260$~GHz, $\delta_{\text{gs}}=50$~GHz are the excited and ground state splittings respectively, $f(t)=\text{sech}(\sigma (t-t_0))$ and $\Delta$ is the two-photon detuning. Note that in order to define the error and ideal Hamiltonians we need only these four transitions. For the SiV$^-$ at zero magnetic fields, we also find that $|\Omega_2^{\text{C1}}|=|\Omega_{1}^{\text{C4}}|$, and $\phi_{\text{C2}}=-\phi_{\text{C}4}$. Since we are further interested for the $R_x$ rotations, we fix $
|\Omega_{1}^{\text{A1}}|=|\Omega_{2}^{\text{A4}}|$. The transformation matrix to the dark-bright frame is given by:
\begin{equation}
\begin{split}
    R&= \frac{\sigma_{11}-\sigma_{14}+\sigma_{41}+\sigma_{44}}{\sqrt{2}}+\sigma_{22}+\sigma_{33}\\&+\sigma_{\text{AA}}+\sigma_{\text{BB}}+\sigma_{\text{CC}}+\sigma_{\text{DD}},
    \end{split}
\end{equation}
which transforms our initial ground states into the dark-bright states (i.e. $|1\rangle \rightarrow |\text{d}\rangle$ and $|4\rangle \rightarrow |\text{b}\rangle$), and the initial Hamiltonian $H$ into $H_{\text{db}}=R HR^\dagger$. Our target Hamiltonian in the db-frame is given by:
\begin{equation}
    H_{0,\text{db}}=(\sigma e^{i\Delta t}\sigma_{\text{bA}}+\text{H.c.})f(t),
\end{equation}
where we have substituted $\sigma=\sqrt{2}|\Omega_{1}^{\text{A1}}|$, and we have defined $|\text{b}\rangle=1/\sqrt{2}(|1\rangle+|4\rangle)$ to be the bright state. Next, we apply one more transformation by going to the interaction picture generated by the ideal Hamiltonian, $H_{0,\text{db}}$, which is given by $U_0$:
\begin{equation}\label{EqD4}
\begin{split}
    U_0&=\sigma_{11}+\sigma_{22}+\sigma_{33}+\sigma_{\text{BB}}+\sigma_{\text{CC}}+\sigma_{\text{DD}}\\&+\cos \theta (\sigma_{44}+\sigma_{\text{AA}})+i\sin\theta (\sigma_{4\text{A}}+\sigma_{\text{A}4}).
    \end{split}
\end{equation}
Note that $U_0$ of Eq.~(\ref{EqD4}) should not be confused with the target gate $U_0$ of Sec.~\ref{SecApp0}. Here, $\theta(t)$ is the integral of the pulse envelope, which for the resonant case reads:
\begin{equation}
\begin{split}
   \theta(t)&=  \int_{0}^{t} \sigma\text{sech}(\sigma (t'-t_0))dt'\\&=2( \tan^{-1}(e^{\sigma(t-t_0)})-\tan^{-1}(e^{-\sigma t_0})).
   \end{split}
\end{equation}
For the off-resonant case, we need to evaluate:
\begin{equation}
\begin{split}
    \theta_\pm(t)&=\int_0^{t}\sigma \text{sech}(\sigma (t'-t_0))e^{\pm i\Delta t'}dt'\\&=\sigma e^{\pm i\Delta t_0} \int_{-t_0}^{t-t_0} e^{\pm i\Delta u}\text{sech}(\sigma u)du.
    \end{split}
\end{equation}
The solution of these indefinite integrals is:
\begin{equation}
    g_{\pm}(u)= {_2}F_1(1,\frac{\sigma \pm i\Delta}{2\sigma},\frac{3\sigma \pm i\Delta}{2\sigma},-e^{2\sigma u})\frac{e^{i[(\sigma \pm \Delta )u \pm  \Delta t_0]}}{\frac{\sigma\pm i\Delta}{2\sigma}},
\end{equation}
where $_2 F_1 (a,b,c,z)$ is the Gauss hypergeometric function. By evaluating this expression in the limits of the integration, we obtain $\theta_{\pm}(t)$. We notice that the two functions $\theta_{\pm}(t)$ are complex conjugates, which simplifies the equations for the off-resonant case. To obtain the set of equation for off-resonant driving, we replace $\theta(t)$ of the resonant case by $\tilde{\theta}(t)=|\theta_+(t)|=|\theta_-(t)|$.

 The error terms in the interaction picture of $H_{0,\text{db}}$, are given by $H_{\text{error}}=U_0(H_{\text{db}}-H_{0,\text{db}})U_0^\dagger$. Using the Gell-Mann basis we now decompose the error terms (where the RWA has been applied) into the operators
\begin{widetext}
\begin{eqnarray}
H_{\text{er},1}^{(1)}&=&\frac{(|\Omega_{1}^{\text{C1}}|+|\Omega_{2}^{\text{C4}}|)\cos(t\delta_{\text{es}})-2|\Omega_{1}^{\text{C4}}|\sin(t\delta_{\text{es}})\sin(\phi_{\text{C}4}+t\delta_{\text{gs}})}{\sqrt{2}}f(t)L_{s,17} \\
H_{\text{er},2}^{(1)}&=& \frac{\cos\theta(t)\cos(\phi_1/2 +\delta_{\text{es}}t)(|\Omega_{1}^{\text{C1}}|-|\Omega_{2}^{\text{C4}}|+2|\Omega_{1}^{\text{C4}}|\cos(\phi_{\text{C}4}+\delta_{\text{gs}}t))}{\sqrt{2}}f(t)L_{s,47}\\
H_{\text{er},3}^{(1)}&=&\frac{\sin\theta(t)\sin(\phi_1/2+\delta_{\text{es}}t)(|\Omega_{1}^{\text{C1}}|-|\Omega_{2}^{\text{C4}}|+2|\Omega_{1}^{\text{C4}}|\cos(\phi_{\text{C}4}+\delta_{\text{gs}}t))}{\sqrt{2}}f(t)L_{s,57}\\
H_{\text{er},4}^{(1)}&=&\frac{(|\Omega_1^{\text{C1}}|+|\Omega_{2}^{\text{C4}}|)\sin(\delta_{\text{es}}t)+2|\Omega_{1}^{\text{C4}}|\cos(\delta_{\text{es}}t)\sin(\phi_{\text{C}4}+\delta_{\text{gs}}t)}{\sqrt{2}}f(t)L_{a,17}\\
H_{\text{er},5}^{(1)}&=&\frac{\cos\theta(t) \sin(\phi_1/2+\delta_{\text{es}}t)(|\Omega_{1}^{\text{C1}}|-|\Omega_{2}^{\text{C4}}|+2|\Omega_{1}^{\text{C4}}|\cos(\phi_{\text{C}4}+\delta_{\text{gs}}t))}{\sqrt{2}}
f(t)L_{a,47}\\
H_{\text{er},6}^{(1)}&=&\frac{\sin\theta(t)\cos(\phi_1/2+\delta_{\text{es}}t)(-|\Omega_{1}^{\text{C1}}|+|\Omega_{2}^{\text{C4}}|-2|\Omega_{1}^{\text{C4}}|\cos(\phi_{\text{C}4}+\delta_{\text{gs}}t))}{\sqrt{2}}f(t)L_{a,57},
\end{eqnarray}
\end{widetext}
 where $L_s$ are the symmetric and $L_a$ the anti-symmetric Gell-Mann operators given by \cite{Gell-Mann}: 
 
 \begin{equation}
     L_{s,jk}=|j\rangle\langle k| + |k\rangle \langle j|
 \end{equation}
\begin{equation}
     L_{a,jk}=-i(|j\rangle\langle k| - |k\rangle \langle j|),
 \end{equation}
 where $1\leq j <k \leq d$, with $d=8$ being the dimension of the Hilbert space.

As we explained in the main text, we need a corrective Hamiltonian that is decomposed into at least the same operators as the error terms.
In other words, starting from the lab frame, we are looking for control pulses that drive the C1 and C4 unwanted transitions. However, it is not a strict requirement that the control pulse drives both error transitions (we will show a counter-example later for the SnV$^-$). 

In the general case, the lab-frame control Hamiltonian for the SiV$^-$ has the form:
\begin{equation}
    W_{\text{lab}}^{(n)}=(\Omega_{1}^{\text{A1}}\sigma_{1\text{A}}+\Omega_{1}^{\text{C1}}\sigma_{1\text{C}}+\Omega_{1}^{\text{C4}}\sigma_{4\text{C}}+\text{H.c.})g^{(n)}(t),
\end{equation}
where  $g^{(n)}(t)=(g_1^{(n)}\cos(\omega_{\text{d}} t)+g_{2}^{(n)}\sin(\omega_{\text{d}} t)$ and $\omega_{\text{d}}$ is the frequency of the control. The amplitudes $g_{1/2}^{(n)}$ are expanded in a Fourier series:
\begin{equation}
    g_{1/2}^{(n)}=\sum_{k}c_{k,1/2}^{(n)}\left(1-\cos\left(\frac{2\pi k t}{T}\right)\right),
\end{equation}
where $T$ is the gate time, $k$ is the order of truncation of the Fourier expansion, and $n$ is the order of truncation of the Magnus series expansion.

We follow the same procedure of transforming our lab frame control Hamiltonian into the interaction picture generated by $H_{0,\text{db}}$. More accurately, we first transform $W_{\text{lab}}^{(n)}(t)$ into the interaction picture via $R_{\text{int}}=\sum_{j} e^{i\omega_jt }|j\rangle\langle j|$ (where $\omega_j$ are the eigenergies), then to the dark-bright frame via $R_{\text{db}}$, and finally into the interaction picture generated by $H_{0,\text{db}}$. After this series of transformations, the decomposition of $W(t)$ in the final frame (and after applying the RWA) yields:
\begin{widetext}
\begin{eqnarray}
W_1&=&\frac{|\Omega_{1}^{\text{A1}}|\sin\theta(-g_2\cos(\phi_1/2+t\Delta_\text{c})+g_1\sin(\phi_1/2+t\Delta_\text{c}))}{\sqrt{2}}L_{s,14} \label{EqD16}\\ 
W_2&=&\frac{|\Omega_{1}^{\text{A1}}|\cos\theta(g_1\cos(\phi_1/2+t\Delta_\text{c})+g_2\sin(\phi_1/2+t\Delta_\text{c}))}{\sqrt{2}}L_{s,15}\\
W_3&=&\frac{|\Omega_1^{\text{C1}}|(g_1\cos(t\bar{\Delta})+g_2\sin(t\bar{\Delta}))+|\Omega_{1}^{\text{C4}}|(g_1\cos(\phi_{\text{C4}}+t\tilde{\Delta}))+g_2\sin(\phi_{\text{C4}}+t\tilde{\Delta}) )}{\sqrt{2}}L_{s,17}\\
W_4&=&\frac{|\Omega_{1}^{\text{A1}}|(g_1\cos(t\Delta_\text{c})+g_2\sin(t\Delta_\text{c}))}{\sqrt{2}}L_{s,45}\\
W_5&=&\frac{\cos\theta(|\Omega_1^{\text{C1}}|(g_1\cos\alpha-g_2\sin\alpha) +|\Omega_{1}^{\text{C4}}|(g_1\cos\beta-g_2\sin\beta)  ) }{\sqrt{2}}L_{s,47}\\
W_6&=&\frac{|\Omega_1^{\text{C1}}|(g_2\cos\alpha+g_1\sin\alpha )+|\Omega_{1}^{\text{C4}}|(g_2\cos\beta+g_1\sin\beta)}{\sqrt{2}}L_{s,57}\\
W_7&=&\frac{|\Omega_{1}^{\text{A1}}|\sin\theta(g_1\cos(\phi_1/2+t\Delta_\text{c})+g_2\sin(\phi_1/2+t\Delta_\text{c}))}{\sqrt{2}}L_{a,14}\\
W_8&=&\frac{|\Omega_{1}^{\text{A1}}|\cos\theta(g_2\cos(\phi_1/2+t\Delta_\text{c})-g_1\sin(\phi_1/2+t\Delta_\text{c}))}{\sqrt{2}}L_{a,15}\\
W_9&=&\frac{|\Omega_1^{\text{C1}}|(g_2\cos(t\bar{\Delta})-g_1\sin(t\bar{\Delta}))+|\Omega_{1}^{\text{C4}}|(-g_2\cos(\phi_{\text{C}4}+t\tilde{\Delta})+g_1\sin(\phi_{\text{C}4}+t\tilde{\Delta}))}{\sqrt{2}}L_{a,17}\\
W_{10}&=&\frac{|\Omega_{1}^{\text{A1}}|\cos(2\theta)(g_2\cos(t\Delta_\text{c})-g_1\sin(t\Delta_\text{c}))}{\sqrt{2}}L_{a,45}\\
W_{11}&=&\frac{\cos\theta( |\Omega_1^{\text{C1}}|  (g_2\cos\alpha +g_1\sin\alpha ) +|\Omega_{1}^{\text{C4}}|( g_2\cos\beta +g_1\sin\beta )  )}{\sqrt{2}}L_{a,47}\\
W_{12}&=&\frac{\sin\theta(|\Omega_1^{\text{C1}}|(-g_1\cos\alpha+g_2\sin\alpha)+|\Omega_{1}^{\text{C4}}|(-g_1\cos\beta+g_2\sin\beta))}{\sqrt{2}}L_{a,57}\\
W_{13}&=&\frac{\sqrt{3}}{4}|\Omega_{1}^{\text{A1}}|\sin(2\theta)(g_2\cos(t\Delta_\text{c})-g_1\sin(t\Delta_\text{c}))L_{33}\\
W_{14}&=&\frac{\sqrt{5}}{4}|\Omega_{1}^{\text{A1}}|\sin(2\theta)(-g_2\cos(t\Delta_\text{c})+g_1\sin(t\Delta_\text{c}))L_{44} \label{EqD29}
\end{eqnarray}
\end{widetext}
with $\bar{\Delta}=\Delta_\text{c}-\delta_{\text{es}}$, $\tilde{\Delta}=\bar{\Delta}+\delta_{\text{gs}}$,  $\alpha=\phi_1/2-t\bar{\Delta}$ and $\beta=\phi_1/2-\phi_{\text{C}4}-t\tilde{\Delta}$. Here $\Delta_\text{c}$ is the detuning of the control measured from the A1 transition, which is fixed to be the same with the detuning of the laser field $\textbf{E}_1$, since we modulate that initial laser. We have also dropped the superscript $n$ in $g_1$, $g_2$ and $W_{i}$, that denotes the order of Magnus truncation. 

The linear system of equations for the first order Magnus expansion is formed as follows:

\begin{widetext}
\begin{equation}
    \left(\begin{array}{ccc|ccc}
    w_{j=1,k=1,l}^{(1)} & \hdots & w_{j=1,k=k_{\mathrm{max}},l}^{(1)} & w_{j=1,k=1,l'}^{(1)} & \hdots & w_{j=1,k=k_{\text{max}},l'}^{(1)} \\
    \vdots & \ddots & \vdots &  \vdots & \ddots & \vdots  \\
    w_{j=j_{\text{max}},k=1,l}^{(1)} & \hdots & w_{j=j_{\text{max}},k=k_{\text{max}},l}^{(1)} &  w_{j=j_{\text{max}},k=1,l'}^{(1)}  & \hdots  & w_{j=j_{\text{max}},k=k_{\text{max}},l'}^{(1)} 
    \end{array}\right)\begin{pmatrix}
    c_{k=1,l}^{(1)}\\
    \vdots\\
    c_{k=k_{\text{max}},l}^{(1)}\\
    c_{k=1,l'}^{(1)}\\
    \vdots\\
    c_{k=k_{\text{max}},l'}^{(1)}
    \end{pmatrix}=\begin{pmatrix}
    h_{\text{err}.,j=1}^{(1)}\\
    \vdots \\
    h_{\text{err}.,j=j_{\text{max}}}^{(1)}
    \end{pmatrix},
\end{equation}
\end{widetext}
with $l=1$ corresponding to $g_1$ and $l'=2$ corresponding to $g_2$. Also, we have defined $h_{\text{err}.j}^{(1)}=-i\int_{0}^{T}dt'
H_{\text{err},j}^{(1)}(t')$ to be the integral of the error term of the $j$-th operator. The components of the first matrix are given by:
\begin{equation}
    w^{(1)}_{j,k,m}=\int_{0}^{T}dt' W_{j,m}(t')\left(1-\cos\left(\frac{2\pi k t'}{T}\right)\right),
\end{equation}
where $W_{j,m}$ corresponds to the coefficient of $g_1$ ($m=l$) or $g_2$ ($m=l'$), for each $j$ operator we decomposed the control into. Since the control is decomposed into more operators than the errors, we set $h_{\text{err},j}=0$, for the components of the error vector where the error Hamiltonian has no decomposition.

Regarding the SnV$^-$, we use $yz$-polarization which leads to two vanishing leakage transitions, i.e. $\Omega_1^{\text{C2}}=0$ and $\Omega_2^{\text{C4}}=0$. In this case, the error terms in the final interaction frame have the decomposition: 

\begin{widetext}
\begin{eqnarray}
H_{\text{er},1}^{(1)}&=&\frac{|\Omega_2^{\text{C2}}|\cos t\bar{\Delta} -|\Omega_1^{\text{C4}}|\cos \beta}{\sqrt{2}}L_{s,27}\\
H_{\text{er,2}}^{(1)}&=&\frac{\cos\theta (|\Omega_2^{\text{C2}}|\cos t\bar{\Delta}  +|\Omega_1^{\text{C4}}|\cos\beta)}{\sqrt{2}}L_{s,47}\\
H_{\text{er,3}}^{(1)}&=&-\frac{\sin\theta  ( |\Omega_2^{\text{C2}}|\sin t\bar{\Delta} +|\Omega_1^{\text{C4}}|\sin \beta  )  }{\sqrt{2}}L_{s,57}\\
H_{\text{er,4}}^{(1)}&=&\frac{-|\Omega_2^{\text{C2}}|\sin t\bar{\Delta} +|\Omega_1^{\text{C4}}|\sin\beta}{\sqrt{2}}L_{a,27}\\
H_{\text{er,5}}^{(1)}&=&-\frac{\cos\theta (|\Omega_2^{\text{C2}}|\sin t\bar{\Delta} +|\Omega_1^{\text{C4}}|\sin \beta)}{\sqrt{2}}L_{a,47}\\
H_{\text{er,6}}^{(1)}&=&-\frac{\sin\theta(|\Omega_2^{\text{C2}}|\cos t\bar{\Delta}+|\Omega_1^{\text{C4}}|\cos\beta) }{\sqrt{2}}L_{a,57},
\end{eqnarray}
\end{widetext}
where $\bar{\Delta}=\Delta-(\delta_{\text{es}}+\delta_{\text{gs}})$ and $\beta=t(\Delta-\delta_{\text{es}}+\delta_{\text{gs}})+2\phi_1$. For the resonant case, $\Delta=0$.

For the control, we start from the following lab-frame corrective Hamiltonian:
\begin{equation}
    W_{\text{lab}}=(\Omega_1^{\text{A2}}\sigma_{\text{2A}}+\Omega_{1}^{\text{C4}}\sigma_{4\text{C}}+\text{H.c.})g^{(n)}(t),
\end{equation}
where we have assumed same polarization as the original laser that drives the A2 transition. Following a similar procedure, we find that the control Hamiltonian in the final frame has the decomposition:

\begin{widetext}
\begin{eqnarray}
W_{1}&=&\frac{|\Omega_1^{\text{A2}}|\sin\theta(-g_2\cos(t\Delta_\text{c})+g_1\sin(t\Delta_\text{c}))}{\sqrt{2}}L_{s,24} \\
W_{2}&=&\frac{|\Omega_1^{\text{A2}}|\cos\theta (g_1\cos(t\Delta_\text{c})+g_2\sin(t\Delta_\text{c}))}{\sqrt{2}}L_{s,25}\\
W_3&=&-\frac{|\Omega_1^{\text{C4}}| (g_1 \cos \beta +g_2\sin\beta)}{\sqrt{2}}L_{s,27}\\
W_4&=&\frac{|\Omega_1^{\text{A2}}|(g_1\cos(t\Delta_\text{c})+g_2\sin(t\Delta_\text{c}))}{\sqrt{2}}L_{s,45}\\
W_5&=&\frac{|\Omega_1^{\text{C4}}|\cos\theta (g_1\cos\beta+g_2\sin\beta)}{\sqrt{2}} L_{s,47}\\
W_6&=&\frac{|\Omega_1^{\text{C4}}|\sin\theta (g_2 \cos\beta -g_1\sin\beta)}{\sqrt{2}} L_{s,57}\\
W_7&=&\frac{|\Omega_1^{\text{A2}}|\sin\theta(g_1\cos(t\Delta_\text{c})+g_2\sin(t\Delta_\text{c}))}{\sqrt{2}}L_{a,24}\\
W_8&=&\frac{|\Omega_1^{\text{A2}}|\cos\theta(g_2\cos(t\Delta_\text{c})-g_1\sin(t\Delta_\text{c}))}{\sqrt{2}}L_{a,25}\\
W_9&=&\frac{|\Omega_1^{\text{C4}}|(-g_2\cos\beta+g_1\sin\beta)}{\sqrt{2}} L_{a,27}\\
W_{10}&=&\frac{|\Omega_1^{\text{A2}}|\cos(2\theta)(g_2\cos(t\Delta_\text{c})-g_1\sin(t\Delta_\text{c}))}{\sqrt{2}}L_{a,45}\\
W_{11}&=&\frac{|\Omega_1^{\text{C4}}|\cos\theta(g_2\cos\beta-g_1\sin\beta) }{\sqrt{2}} L_{a,47}\\
W_{12}&=&-\frac{|\Omega_1^{\text{C4}}|(g_1\cos\beta+g_2\sin\beta)}{\sqrt{2}} L_{a,57}\\
W_{13}&=&\frac{\sqrt{3}}{4}|\Omega_1^{\text{A2}}|\sin(2\theta)(g_2\cos(t\Delta_\text{c})-g_1\sin(t\Delta_\text{c}))L_{33}\\
W_{14}&=&\frac{\sqrt{5}}{4}|\Omega_1^{\text{A2}}|\sin(2\theta)(-g_2\cos(t\Delta_\text{c})+g_1\sin(t\Delta_\text{c}))L_{44},
\end{eqnarray}
\end{widetext}
where we have defined $\beta=t(\Delta-\delta_{\text{es}}+\delta_{\text{gs}})$. The detuning of the control is set equal to the two-photon detuning (same frequency as the $\textbf{E}_1$ laser field), i.e. $\Delta_c=\Delta$. Notice that even though we started with a control pulse that does not have access to the error transition C2, in the final interaction frame the linear system of equations is well-defined, as for each error decomposition term, there is a corresponding control decomposition.

Even though the controls are obtained in the interaction frame generated by the ideal Hamiltonian, the fidelity of the Magnus scheme in the main text is evaluated in the initial dark-bright (interaction) frame.

\section{Pulse corrections obtained via the DRAG method \label{SecApp6}}

Here, we provide further details regarding the derivation of the control pulses based on the DRAG method. First, we briefly highlight the strategy for deriving the controls. Following the procedure of Ref. \cite{PhysRevA.83.012308}, we start by transforming our Hamiltonian that also includes the error terms in the rotating frame. We further mention that for the derivation of the corrections, we consider only the subspace composed of the $\{|1\rangle,|4\rangle,|\text{A}\rangle,|\text{C}\rangle\}$ ($\{|2\rangle,|4\rangle,|\text{A}\rangle,|\text{C}\rangle\}$) states for the SiV$^-$  (for the SnV$^-$).

Regarding the SiV$^-$ system, the Hamiltonian for this reduced subspace in the lab frame reads:
\begin{equation}\label{EqE1}
    H=H_{\text{lab},1}+H_{\text{lab},2}+H_0,
\end{equation}
where $H_0=\text{diag}[\omega_1,\omega_1+\delta_{\text{gs}},\omega_A,\omega_A+\delta_{\text{es}}]$, with $\delta_{\text{gs}}$ and $\delta_{\text{es}}$ being the ground and excited states splittings and $\omega_1$, $\omega_A$ being the eigen-energies of $|1\rangle$ and $|\text{A}\rangle$ respectively. Also, $H_{\text{lab},1}$ and $H_{\text{lab},2}$ are given by:

\begin{widetext}
\begin{equation} 
\begin{split}
H_{\text{lab},1}^{(n)}=\Big(\Omega_{1}^{(n)}\cos(\omega_{\text{d1}}t)\Big[e^{i\phi_{\text{A1}}}\sigma_{\text{1A}}+\lambda_1\sigma_{\text{1C}}+\lambda_{12}e^{i\phi_{C4}}\sigma_{\text{4C}}\Big]&+\Omega_{2}^{(n)}\cos(\omega_{\text{d2}}t)\Big[e^{i\phi_{\text{A4}}}\sigma_{\text{4A}}-\lambda_{2}\sigma_{\text{4C}}+\lambda_{21}e^{i\phi_{\text{C1}}}\sigma_{\text{1C}}]\Big)f(t)\\&+\text{H.c.},
\end{split}
\end{equation}

\begin{equation} 
\begin{split}
H_{\text{lab},2}^{(n)}=\Big(\bar{\Omega}_{1}^{(n)}\sin(\omega_{\text{d1}}t)\Big[e^{i\phi_{\text{A1}}}\sigma_{\text{1A}}+\lambda_1\sigma_{\text{1C}}+\lambda_{12}e^{i\phi_{C4}}\sigma_{\text{4C}}\Big]&+\bar{\Omega}_{2}^{(n)}\sin(\omega_{\text{d2}}t)\Big[e^{i\phi_{\text{A4}}}\sigma_{\text{4A}}-\lambda_{2}\sigma_{\text{4C}}+\lambda_{21}e^{i\phi_{\text{C1}}}\sigma_{\text{1C}}]\Big)f(t)\\&+\text{H.c.}~,
\end{split}
\end{equation}
\end{widetext}
with $f(t)=\text{sech}(\sigma(t-t_0))$ and $\omega_{\text{d}1}$, $\omega_{\text{d}2}$ the laser frequencies. The fields $\bar{\textbf{E}}_1$ and $\bar{\textbf{E}}_2$ are $\pi/2$-shifted compared to $\textbf{E}_1$ and \textbf{E}$_2$. Starting from Eq.~(\ref{EqE1}) we perform the transformation

\begin{equation}
U_{\text{rot}}=\text{diag}[e^{i\omega_{1}t}, e^{i(\omega_1+\omega_{\text{d1}}-\omega_{\text{d2}})t},e^{i(\omega_1+\omega_{\text{d1}})t},e^{i(\omega_1+\omega_{\text{d1}})t}],
\end{equation}
which leads to the rotating frame Hamiltonian, as well as the transformation
\begin{equation}
    U_\phi=\text{diag}[e^{-i\phi_{\text{A1}}},e^{-i\phi_{\text{A1}}},1,e^{-i\phi_{\text{A1}}}].
\end{equation}

Notice that the transformation $U_\phi$ removes the complex part $e^{i\phi_{\text{A1}}}$ from the Rabi frequency corresponding to the A1 as well as A4 transitions, since we fix the Rabi frequencies to be equal a priori to satisfy the db transformation for $R_x$ gates. At this step, our rotating frame Hamiltonian reads:
\begin{widetext}
\begin{equation}
\begin{split}
H_{\text{rot}}&=\frac{1}{2}\Big[\Big(\Omega_{1}\sigma_{1\text{A}}+\Omega_{2}\sigma_{4\text{A}}+(\lambda_1\Omega_1 +e^{-it(\delta_{\text{gs}}+\phi_{\text{C}4})}\lambda_{21}\Omega_2)\sigma_{1\text{C}}   + (e^{it(\delta_{\text{gs}}+\phi_{\text{C}4})}\lambda_{12}\Omega_1-\lambda_2\Omega_2)\sigma_{4\text{C}} \Big) 
+ \text{H.c.}\Big] 
\\&+
\frac{-i}{2}\Big[\Big(\bar{\Omega}_{1}\sigma_{1\text{A}}+\bar{\Omega}_{2}\sigma_{4\text{A}}+(\lambda_1\bar{\Omega}_1 +e^{-it(\delta_{\text{gs}}+\phi_{\text{C}4})}\lambda_{21}\bar{\Omega}_2)\sigma_{1\text{C}}   + (e^{it(\delta_{\text{gs}}+\phi_{\text{C}4})}\lambda_{12}\bar{\Omega}_1-\lambda_2\bar{\Omega}_2)\sigma_{4\text{C}} \Big) 
+ \text{H.c.}\Big]
\\&-
\Delta\sigma_{\text{AA}}+(\delta_{\text{es}}-\Delta)\sigma_{\text{CC}}.
\end{split}
\end{equation}
\end{widetext}
For clarity, we mention that we have defined $\lambda_1=|\Omega_1^{\text{C1}}|/|\Omega_1|$, $\lambda_{12}=|\Omega_{1}^{\text{C4}}|/|\Omega_1|$, $\lambda_{2}=|\Omega_{2}^{\text{C4}}|/|\Omega_2|$ and $\lambda_{21}=|\Omega_2^{\text{C1}}|/|\Omega_2|$, where the subscripts $k=\{1,2\}$ in $\Omega_k^{ij}$ correspond to the lasers by which the error transitions are driven by. 
Finally, in order to go to the db-frame we apply the transformation:
\begin{equation}
    R_{\text{db}}=\frac{1}{\sqrt{2}}(\sigma_{11}-\sigma_{14}+\sigma_{41}+\sigma_{44})+\sigma_{\text{AA}}+\sigma_{\text{CC}}.
\end{equation}
To decouple the dark state from the excited, we further set $\Omega_1=\Omega_2$ and $\bar{\Omega}_1=\bar{\Omega}_2$. Thus, the dark-bright (rotating) Hamiltonian reads:
\begin{widetext}
\begin{equation}
\begin{split}
    H_{\text{db}}=\frac{1}{\sqrt{2}}\Big(
    &-\frac{(e^{it(\delta_{\text{gs}}+\phi_{\text{C}4})}\lambda_{12}-(\lambda_1+\lambda_2)-\lambda_{21}e^{-it(\delta_{\text{gs}}+\phi_{\text{C}4})} )(\Omega_1-i\bar{\Omega}_1)  }{2}\sigma_{\text{dC}}+(\Omega_1-i\bar{\Omega}_1)\sigma_{\text{bA}} 
    \\&+ 
    \frac{(e^{it(\delta_{\text{gs}}+\phi_{\text{C}4})}\lambda_{12}+(\lambda_1-\lambda_2)+\lambda_{21}e^{-it(\delta_{\text{gs}}+\phi_{\text{C}4})})(\Omega_1-i\bar{\Omega}_1)}{2}\sigma_{\text{bC}}+\text{H.c.}
    \Big)
    \\&-\Delta \sigma_{\text{AA}}+(-\Delta+\delta_{\text{es}})\sigma_{\text{CC}}.
    \end{split}
\end{equation}
\end{widetext}

The leakage subspace $|\text{C}\rangle$ is off-resonant from the remaining Hamiltonian by an energy cost $\delta_{\text{es}}$. Effectively, this allows us to perform an expansion of the control fields in the parameter $\epsilon=1/(T\delta_{\text{es}})$. More analytically, according to Ref.~\cite{PhysRevA.83.012308}, by multiplying $H_{\text{db}}$ by the gate time we convert it to the dimensionless form:
\begin{equation}
    \tilde{H}_{\text{db}}=\frac{1}{\epsilon}H_0+\sum_{n=0}^\infty \epsilon^n \tilde{H}_{\text{db}}^{(n)}(t),
\end{equation}
with $H_0=\text{diag}[0,0,0,1]$ and $\tilde{H}_{\text{db}}^{(n)}(t)$ given by:
\begin{widetext}
\begin{equation}\label{EqE8}
\begin{split}
    \tilde{H}_{\text{db}}^{(n)}(t)=\frac{1}{\sqrt{2}}\Big(
    &-\frac{(e^{it(\delta_{\text{gs}}+\phi_{\text{C}4})}\lambda_{12}-(\lambda_1+\lambda_2)-\lambda_{21}e^{-it(\delta_{\text{gs}}+\phi_{\text{C}4})} )(\Omega_1^{(n)}-i\bar{\Omega}_1^{(n)})  }{2}\sigma_{\text{dC}}+(\Omega_1^{(n)}-i\bar{\Omega}_1^{(n)})\sigma_{\text{bA}} 
    \\&+ 
    \frac{(e^{it(\delta_{\text{gs}}+\phi_{\text{C}4})}\lambda_{12}+(\lambda_1-\lambda_2)+\lambda_{21}e^{-it(\delta_{\text{gs}}+\phi_{\text{C}4})})(\Omega_1^{(n)}-i\bar{\Omega}_1^{(n)})}{2}\sigma_{\text{bC}}+\text{H.c.}
    \Big)
    \\&-\Delta \sigma_{\text{AA}}+(-\Delta+\delta_{\text{es}})\sigma_{\text{CC}}.
    \end{split}
\end{equation}
\end{widetext}
Note that now in Eq.~(\ref{EqE8}), the control fields $\Omega_{k}^{(n)}$ and $\bar{\Omega}_{k}^{(n)}$, as well as the detuning, $\Delta^{(n)}$, should be understood as dimensionless. The next step is to satisfy the target constraints that will allow us to implement the ideal Hamiltonian of Eq.~(\ref{Eq25}) of Sec.~\ref{Sec4b}, as well as the decoupling constraints that will suppress the leakage to the $|\text{C}\rangle$ subspace. These constraints are imposed in the DRAG frame, which as we mentioned in the main text is generated by the Hermitean operator $S(t)$, via $A=e^{-iS(t)}$. To this end, the operator $S(t)$ is expanded in power series in $\epsilon$, as $S(t)=\sum_{n=1}\epsilon^n S^{(n)}(t)$, with $S^{(n)}(t)$ given by:
\begin{equation}
    S^{(n)}(t)=\sum_{j=1}s^{(n)}_{z,j}\sigma_{jj}+\sum_{j<k}s^{(n)}_{x,jk}\sigma_{x,jk}+
    \sum_{j<k}s^{(n)}_{y,jk}\sigma_{y,jk}.
\end{equation}

As a result, the decoupling and target constraints can be solved iteratively in a consistent manner, and the set of equations for the $n$-th order can be found in the Appendix of Ref.~\cite{PhysRevA.83.012308}. 
For transparency, we highlight how we solve the constraints and provide the equations for the corrective modulations. 

The first step is to define the target Hamiltonian, which as given in the main text reads:
\begin{equation}
    H_{\text{target}}=\frac{h_x^{(0)}}{2}\sigma_{x,\text{be}}+h_z^{(0)}(\sigma_{\text{bb}}-\sigma_{\text{ee}}).
\end{equation}
By satisfying the zero-th order constraints we ensure that $H_{\text{D}}^{(0)}=H_{\text{db},0}$, where $H_{\text{db},0}$ is the ideal Hamiltonian:
\begin{equation}
    H_{\text{db},0}=(\Omega_{eff}f(t)\sigma_{x,\text{be}}+\text{H.c.})-\Delta\sigma_{\text{ee}}.
\end{equation}
Effectively, this means that to the zero-th order, the target gate is the same in both frames. At the same time, satisfying the zero-th order constraints implies that certain $S^{(1)}(t)$ elements need to be restricted; these correspond to $S^{(1)}_{k,\text{dC}}$, $S^{(1)}_{k,\text{bC}}$, $S^{(1)}_{k,\text{AC}}$, with $k=\{x,y\}$. This leaves the parameters $S^{(1)}_{z,j}$ (with $j=\{\text{d, b, A, C}\}$), $S^{(1)}_{k,\text{db}}$, $S^{(1)}_{k,\text{dA}}$ and $S^{(1)}_{k,\text{bA}}$ free. We set all $S^{(1)}_{z,j}=0$, as well as $S^{(1)}_{k,\text{db}}=S^{(1)}_{k,\text{dA}}=0=S^{(1)}_{x,\text{bA}}$. This choice satisfies the boundary conditions for the frame transformation $A(t)$, and allows us to obtain the corrective fields by $S^{(1)}_{y,\text{bA}}(t)$ via the first order target constraints. In particular, for $\Delta^{(1)}=0$, the target condition:
\begin{equation}
    \text{Tr}[H_{\text{D}}^{(1)}(\sigma_{\text{bb}}-\sigma_{\text{ee}})]=0,
\end{equation}
gives the following solution for $S^{(1)}_{y,\text{bA}}(t)$:
\begin{equation}
    S^{(1)}_{y,\text{bA}}(t)=\frac{\Omega_1^{(0)}(\lambda_1-\lambda_2 + 2\lambda_{12} \cos(t\delta_{\text{gs}}+\phi_{\text{C}4}))^2f(t)}{8\sqrt{2}\delta_{\text{es}}},
\end{equation}
where we have also set $\lambda_{12}=\lambda_{21}$, which arises from the polarization definitions we have used. Then, from the target constraints:
\begin{eqnarray}
h_x^{(1)}&=&\text{Tr}[H_{\text{D}}^{(n)}\sigma_{x,\text{be}}]=0,\\
h_y^{(1)}&=&\text{Tr}[H_{\text{D}}^{(n)}\sigma_{y,\text{be}}]=0,
\end{eqnarray}
we solve for $\Omega_1^{(1)}$ and $\bar{\Omega}_1^{(1)}$, which depend on $S^{(1)}_{y,\text{bA}}(t)$. The expressions for the pulse corrections for the SiV$^-$ are:

\begin{equation}
    \Omega_1^{(1)}=\Omega_2^{(1)}=\frac{\Delta \Omega_{1}^{(0)}(\lambda_1-\lambda_2+2\lambda_1\lambda_2\cos(t\delta_{\text{gs}}+\phi_{\text{C4}}))^2}{16\delta_{\text{es}}},
\end{equation}
\begin{equation}
    \bar{\Omega}_1^{(1)}=\bar{\Omega}_2^{(1)}=\frac{\Omega_1^{(0)}(\lambda_1-\lambda_2+2\lambda_{12}\cos(t\delta_{\text{gs}}+\phi_{\text{C4}}))}{16\delta_{\text{es}}}A(t),
\end{equation}
where $A(t)$ is given by:
\begin{equation}
\begin{split}
    A(t)=\Big(&-4\delta_{\text{gs}}\lambda_{12}\sin(t\delta_{\text{gs}}+\phi_{\text{C4}})\\&+\left(\lambda_1-\lambda_2+2\lambda_{12}\cos(t\delta_{\text{gs}}+\phi_{\text{C4}})\right)\frac{\dot{f}(t)}{f(t)} \Big)
    \end{split}
\end{equation}

Lastly, we follow a similar procedure for the SnV$^-$, and we find that the corrections $\Omega_1^{(1)}$ and $\bar{\Omega}_1^{(1)}$ are:

\begin{equation}
\Omega_1^{(1)}=\Omega_2^{(1)}=\frac{\Omega_1^{(0)}\Delta ( (\lambda_1+\lambda_2)^2+2\lambda_{21}^2(1-\cos(2\delta_{\text{gs}}t))  )}{16\delta_{\text{es}}}    
\end{equation}
\begin{equation}
    \bar{\Omega}_1^{(1)}=\bar{\Omega}_2^{(1)}=\frac{\lambda_{21}^2\Omega_1^{(0)}\sin(t\delta_{\text{gs}})(2 \delta_{\text{gs}}\cos(t\delta_{\text{gs}})+\frac{\dot{f}(t)}{f(t)}\sin(t\delta_{\text{gs}}))}    {4\delta_{\text{es}}},
\end{equation}
where $\lambda_{21}=\Omega_2^{\text{C2}}/\Omega_1^{(0)}$.

\end{document}